\documentclass[12pt, draftcls, letterpaper, onecolumn, journal]{IEEEtran}
\usepackage{cite,subfigure,psfrag,graphicx,amssymb,color}
\usepackage{graphicx,amssymb}
\usepackage[tbtags]{amsmath}
\usepackage{times,amsmath}
\usepackage{subfigure}
\usepackage{epsfig}
\usepackage{amssymb}
\usepackage{hyperref}
\usepackage{color}
\usepackage{subfigure}

\begin{document}
\title{\vspace{0cm}\LARGE\bf Cooperative Precoding/Resource
Allocation Games 
under Spectral Mask and Total Power Constraints}

\author{{Jie Gao, \IEEEmembership{Student Member, IEEE},
Sergiy A. Vorobyov, \IEEEmembership{Senior Member, IEEE},\\
and Hai Jiang, \IEEEmembership{Member, IEEE}}
\thanks{The authors are with the Department of Electrical and Computer
Engineering, University of Alberta, Edmonton, Alberta, Canada. The
contacting emails are \{jgao3, vorobyov,
hai.jiang\}@ece.ualberta.ca.}
\thanks{{\bf Corresponding author:} Sergiy A. Vorobyov, Dept. of \
Electrical and Computer Engineering, University of Alberta,
9107-116~St., Edmonton, Alberta, T6G 2V4, Canada; Phone:
+1 (780) 492 9702, Fax: +1 (780) 492 1811.}
\thanks{This work was supported in parts by research grants from the
Natural Science and Engineering Research Council (NSERC) of Canada
and Alberta Ingenuity New Faculty Award.} }

\maketitle

\vspace{-1.5cm}
\begin{abstract}
The use of orthogonal signaling \ schemes such as time-,
frequency-, or \ code-division \ multiplexing (T-, F-, CDM) in
multi-user systems allows for power-efficient simple receivers. It
is shown in this paper that by using orthogonal signaling on
frequency selective fading channels, the cooperative Nash
bargaining (NB)-based precoding games for multi-user systems,
which aim at maximizing the information rates of all users, are
simplified to the corresponding cooperative resource allocation
games. The latter provides additional practically desired
simplifications to transmitter design and significantly reduces
the overhead during user cooperation. The complexity of the
corresponding precoding/resource allocation games, however,
depends on the constraints imposed on the users. If only spectral
mask constraints are present, the corresponding cooperative NB
problem can be formulated as a convex optimization problem and
solved efficiently in a distributed manner using dual
decomposition based algorithm. However, the NB problem is
non-convex if total power constraints are also imposed on the
users. In this case, the complexity associate with finding the NB
solution is unacceptably high. Therefore, the multi-user systems
are categorized into bandwidth- and power-dominant based on a
bottleneck resource, and different manners of cooperation are
developed for each type of systems for the case of two-users. Such
classification guarantees that the solution obtained in each case
is Pareto-optimal and actually can be identical to the optimal
solution, while the complexity is significantly reduced.
Simulation results demonstrate the efficiency of the proposed
cooperative precoding/resource allocation strategies and the
reduced complexity of the proposed algorithms.
\end{abstract}

\vspace{-0.2cm} {\bf Index Terms:} Cooperative games, multi-user
systems, Nash bargaining, dual decomposition, Pareto-optimality,
spectral mask constraints, total power constraints.

\section{Introduction}\label{s:int}
In multi-user systems, all users compete for resources and can
cause interference to each other. This makes it impossible for any
user to gain more profit without harming other users. The
traditional information-theoretic studies of multi-user systems
are mainly focused on finding the corresponding rate regions and
do not advise how to actually achieve the best rates for all users
simultaneously (see \cite{informth2}--\cite{informth5} and the
references therein). It is, however, evident that the performance
of multi-user systems depends on the balance among the users in
the competition for resources. Moreover, the points in the
achievable rate region are not all stable, or even feasible if the
selfish nature of the users is taken into account. Indeed, it is
reasonable to assume that all users will compete for the maximum
achievable benefits at all times, which may render difficulties to
the implementation of any prescribed regulations against the
selfishness of users. For example, although an outcome
corresponding to the case when one user is forced to sacrifice its
performance for the benefit of other users can be theoretically
justified, it is hard to make sure in practice that the sacrificed
user will not deviate from the regulation which is unfair for him.

Recently, game theory has been recognized as an appropriate tool
for studying multi-user systems \cite{Poor}--\cite{MISOgame}. It
studies the actions of decision makers (players, here wireless
users) with conflicting objectives, and predicts the users'
decisions on future actions (strategies) and the outcome of the
game. If users compete with each other, the existence of ``stable"
outcomes, corresponding to the so-called {\it equilibria}, can be
analyzed \cite{vaderSchaar}--\cite{Gesualdo2}. On the other hand,
if there is a voluntary cooperation among users, the extra
benefits for all users can be obtained. The corresponding games
are called {\it cooperative} games and one of the most popular
approaches developed for cooperative games is the Nash bargaining
(NB) approach \cite{coop_game}.

Although the use of cooperative game theory to recourse allocation
in multi-user wireless systems is a recent research topic, there
are some results available. A two-user power allocation game on a
flat fading channel (FFC) is investigated in \cite{Leshem}. It is
argued that certain points in the utility space of the game (i.e.,
the information-theoretic rate region of the multi-user system)
are not achievable from a game-theoretic perspective. It is also
shown that the NB solution based on time division multiplexing
(TDM) increases the benefits of all users as compared to the
non-cooperative Nash equilibrium (NE) solution. The study is
extended in \cite{Leshem2} to $N$-user systems with frequency
selective fading channels (FSFCs) for the case when only spectral
mask constraints (SMCs) are imposed on the users. The NB solution
is derived based on joint TDM/frequency division multiplexing
(TDM/FDM) scheme. Unlike the FFC case, the allocation of frequency
bins becomes a major problem on the FSFC. A more complex resource
allocation game on the FSFC with only total power constraints
(TPCs) limiting the total transmission power of each user is
considered in \cite{Leshem4}. A water-filling based algorithm is
proposed to search for the NB solution in a two-user version of
the game. The proposed algorithm bargains in many different convex
subspaces of the original utility space and obtains one NB
solution in each subspace. Then, the NB solution with the largest
outcome is selected. However, the TPCs render the complexity of
the algorithm high.

One more application area of cooperative game theory is
beamforming in multiple-input single-output (MISO) systems
\cite{Larsson}--\cite{MISOgame}. A two-user game on interference
channel is investigated in \cite{Larsson}, where user strategies
are defined as the choices of beamforming vectors. The superiority
of the cooperative NB solution over the non-cooperative NE
solution is demonstrated, and some special points such as sum-rate
and zero-forcing points are shown to be unstable from a
game-theoretic viewpoint. Kalai-Smorodinsky-type solutions of
cooperative beamforming games are further derived in
\cite{ksMISO}. For the games on two-user MISO systems, it is also
shown in \cite{MISOnezf} and \cite{MISOgame} that any
Pareto-optimal point in the game's utility space can be realized
through a certain balance between competition and cooperation
among the users.

Game theory has been also used for precoding design. A
non-cooperative precoding game is analyzed in \cite{Gesualdo}
under SMCS and TPCs, in which a multi-user FSFC is considered and
the optimal precoding matrices are derived based on the NE. It is
shown that the matrix-valued precoding games boil down to
equivalent vector-valued power allocation games, and the resulted
precoding matrices adopt a diagonal structure.\footnote{The
precoding matrices were mistakenly expressed in \cite{Gesualdo} as
a product of the inverse fast Fourier transform (IFFT) matrices
and power allocation diagonal matrices, while they should be
expressed only as power allocation diagonal matrices.} The
existence and uniqueness of the NE is guaranteed if the
communication links are sufficiently far away from each other, and
the NE is more efficient when the interference power is relatively
low as compared to the noise power. Although non-cooperative games
do not coordinate users and, therefore, allow for low-complexity
and distributed solutions, they often lead to quite inefficient
results for all users due to the lack of coordination.

In this paper, we develop cooperative NB-based precoding
strategies for the multi-user wireless systems using cooperative
game theory.\footnote{Some preliminary results without proofs have
been reported in \cite{Precoding_game}, \cite{MultiPrecoding}, and
submitted \cite{POresalloc}.} The main contribution of this paper
is threefold. First, it is shown that cooperative precoding games
boil down to cooperative resource allocation games under
orthogonal signaling set up, that is, the TDM-based cooperation
among users for FFCs or the TDM/FDM-based cooperation for FSFCs.
The precoding matrices adopt a strictly diagonal structure in
these cases. Second, we show that the process of bargaining among
users can be physically realized in a distributed and efficient
manner with very low information overhead if only SMCs are imposed
on the users. Third, efficient algorithms for the
precoding/resourse allocation games are developed for the case
when both SMCs and TPCs are imposed on the users. Although the
bargaining problem appears to be non-convex, efficient algorithms
are designed based on a proposed classification of the multi-user
systems into bandwidth- and power-dominant. Then, different
manners of cooperation are developed for each type of the systems.
Such classification guarantees that the solution obtained for each
type of the systems is Pareto-optimal and actually can be
identical to the optimal solution. Moreover, the complexity is
significantly reduced as compared to the complexity required for
solving the original problem using exact algorithms.

The rest of this paper is organized as follows. The signal model
is introduced and the cooperative precoding/resource allocation
game is formulated in Section~\ref{s:sm_long}. The
precoding/resource allocation strategies for cooperative games
with SMCs are studied in Section~\ref{s:smc}.
Section~\ref{s:smctpc} deals with the two-user games with both
SMCs and TPCs. Section~\ref{s:sr} demonstrates our simulation
results. It is followed by Section~\ref{s:con} that concludes the
paper. All proofs for
Sections~\ref{s:sm_long},~\ref{s:smc},~and~\ref{s:smctpc} are
summarized in Appendices~A,~B,~and~C, respectively. This paper is
reproducible research \cite{Covasevic} and the software needed to
generate the numerical results can be obtained from {\tt
www.ece.ualberta.ca/$\sim$vorobyov/} {\tt ProgNB.zip}.

\section{System model and Precoding/Resourse Allocation Game
Formulation} \label{s:sm_long}

\subsection{System model} \label{s:sysmodel}
Consider an $M$-user wireless system in which all users transmit
on the same wideband FSFC with channel length $L$ where $L$
depends on the channel delay spread and the signal symbol duration
\cite{OFDMbook}.\footnote{FFC can be viewed as FSFC with $L=1$.}
Assuming block transmission with block length $N$ for all users,
the general signal model for user~$i$ can be written as
\begin{equation}\label{e:Sysmodelfsc}
\begin{gathered}
\textbf{y}_i=\textbf{G}_{i}\textbf{H}_{ii}\textbf{F}_i\textbf{s}_i
+ \textbf{G}_{i} \sum\limits_{j=1,j\neq i}^{M}{\textbf{H}_{ji}
\textbf{F}_j\textbf{s}_j}+\textbf{G}_{i}\textbf{n}_i
\end{gathered}
\end{equation}
where $\textbf{s}_i$ is the $N\times1$ information symbol block of
user~$i$, $\textbf{F}_i$ is the $N\times N$ precoding matrix of
user~$i$, $\textbf{G}_i$ is the $N\times N$ decoding matrix of
user~$i$, ${\textbf{H}_{ji}}$ is the $N\times N$ channel matrix
between users~$j$~and~$i$, ${\textbf{n}_i}$ is the $N\times1$
zero-mean additive white Gaussian noise vector with covariance
$E\{\textbf{n}_i\textbf{n}_i^H\} = \sigma_i^2 \textbf{I}$,
$\sigma_i^2$ is the variance of $\textbf{s}_i$, and $( \cdot )^H$,
$E \{ \cdot \}$, and $\textbf{I}$ stand for the Hermitian
transpose, expectation operation, and identity matrix,
respectively. The information symbols are assumed to have
unit-energy and be uncorrelated to each other and to noise, i.e.,
$E\{\textbf{s}_i\textbf{s}_i^H\}=\textbf{I}$ and
$E\{\textbf{s}_i\textbf{n}_i^H\}=\textbf{0}$, where $\textbf{0}$
denotes the matrix of zeros.

In order to decompose a wideband FSFC to flat fading frequency
bins, orthogonal frequency division multiplexing (OFDM) is
adopted. Specifically, assuming that the block length $N$ is
larger than the channel length $L$, introducing cyclic prefix
(CP), and performing IFFT and fast Fourier transform (FFT) at the
transmitter and receiver sides, respectively, the signal model can
be written as \cite{Gesualdo}, \cite{usingOFDM2}, \cite{OFDMbook}
\begin{equation}\label{e:Sysmodelfsc}
\begin{gathered}
\textbf{y}_i = \textbf{G}_{i} \mathbf{\Phi}_{ii} \textbf{F}_i
\textbf{s}_i + \textbf{G}_{i} \tilde{ \textbf{n}}_i \\
\end{gathered}
\end{equation}
where $\tilde{\textbf{n}}_i = \sum\limits_{j=1,j\neq i}^{M}
\mathbf{\Phi}_{ji} \textbf{F}_j \textbf{s}_j + \textbf{D}
\textbf{n}_i$ is the $N\times 1$ interference plus noise vector of
user $i$ before the decoder, $\textbf{D}$ is the FFT matrix,
$\mathbf{\Phi}_{ji}$ is the $N\times N$ diagonalized channel
matrix between users $j$ and $i$ with its $k$th element being the
sampled frequency response of the $k$th frequency bin. Both the
desired communication channel $\textbf{H}_{ii}$ and the
interference channels $\textbf{H}_{ji} \; (\forall j, \, j\neq i)$
are diagonalized due to the CP insertion and the multiplication by
matrices $\textbf{D}^H$ and $\textbf{D}$ at the transmitter and
receiver sides, respectively.

Considering the general case when all users treat the interference
as additive noise, the noise covariance for user $i$ before the
decoder can be expressed as
\begin{equation}\label{e:initem}
\textbf{R}_{-i} = E\{\tilde{\textbf{n}}_i \tilde{\textbf{n}}_i^H\}
= \sigma_i^2 \textbf{I} + \sum\limits_{j=1,j \neq i}^{M}
\mathbf{\Phi}_{ji} \textbf{F}_j \textbf{F}_j^{H}
\mathbf{\Phi}_{ji}^H.
\end{equation}
Then, the Wiener filter is the optimal capacity-lossless linear receiver
\cite{precoder1}, \cite{precoder2}. Thus, the decoding matrix
$\textbf{G}_i$ can be found as
\begin{equation}\label{e:G}
\textbf{G}_i = \textbf{F}_i^H \textbf{H}_{ii}^H (\textbf{H}_{ii}
\textbf{F}_i \textbf{F}_i^H \textbf{H}_{ii}^H +
\textbf{R}_{-i})^{-1}
\end{equation}
and the mutual information (information rate) that user $i$ can
achieve is expressed as \cite{Capacity}
\begin{equation}\label{e:rate}
R_i(\textbf{F}_i,\textbf{F}_{-i}) = \log \left( \det \left(
\textbf{I} + \textbf{F}_{i}^{H} \mathbf{\Phi}_{ii}^{H}
\textbf{R}_{-i}^{-1} \mathbf{\Phi}_{ii}\textbf{F}_{i} \right)
\right)
\end{equation}
where $\textbf{F}_{-i}$ is the set of the precoding matrices of
all users except user $i$ and $\det(\cdot)$ denotes the
determinant.

In practice, all users attempt to maximize their information rates
under certain power constrains. For the case of FSFC, SMCs are
usually considered to limit the powers that the users can allocate
on different frequency bin. These power limits are denoted as
${p^{\max}_i(k)}\; (\forall i\in\Omega_M,\forall k\in\Omega_N$)
where $\Omega_M = \{1, \cdots, M\}$ is the set of user indexes,
$\Omega_N = =\{1, \cdots, N\}$ is the set of frequency bin
indexes. Although SMCs also bound the total power by the value
$\sum\limits_{k\in\Omega_N} p_i^{\max}(k)$ for user $i$, such
bound may be loose compared to possibly imposed total power limit
$P^{\max}_i$. Thus, TPCs may also be needed. The aforementioned
SMCs and TPCs can be mathematically expressed, correspondingly, as
\begin{eqnarray}\label{e:smc}
E\{|[\textbf{F}_i \textbf{s}_i]_k|^2\} \!\!\!&=&\!\!\!
[\textbf{F}_i \textbf{F}_i^H]_{kk} \leq {p_i^{\max}(k)}, \;
\forall i \in \Omega_M, \, \forall k \in \Omega_N \\
\label{e:tpc} E\{\|\textbf{F}_i \textbf{s}_i\|^2\} \!\!\!&=&\!\!\!
{\rm Tr}\{\textbf{F}_i\textbf{F}_i^H\} \leq P^{\max}_i, \; \forall
i \in \Omega_M
\end{eqnarray}
where $[\cdot]_{kk}$ and ${\rm Tr} \{ \cdot \}$ denotes the $k$th
diagonal element and the trace of a square matrix.

\subsection{Cooperative Precoding/Resourse Allocation Game}
Considering the wireless users as players, the choices of
precoding matrices as user strategies, and the corresponding
information rates $R_i$'s as user utilities, the game model of the
precoding problem can be written as
\begin{equation}\label{e:Sysmodelgame}
\Gamma=\Big\{\Omega_M,\{\textbf{F}_i \, | \, {i\in \Omega_M}\},
\{R_i (\textbf{F}_i,\textbf{F}_{-i}) \, | \, {i\in \Omega_M\}}
\Big\}.
\end{equation}
In the non-cooperative case, when the game players (wireless
users) do not collaborate, the NE is a stable strategy combination
of the game that satisfies
\begin{equation}
R_i(\textbf{F}_i^{\text{NE}}, \textbf{F}_{-i}^{\text{NE}}) \geq
R_i(\textbf{F}_i^\prime, \textbf{F}_{-i}^{\text{NE}}),  \quad
\forall \textbf{F}_i^\prime, \, \forall i \in \Omega_M
\end{equation}
where $\textbf{F}_i^{\text{NE}}$ is the precoding strategy of user
$i$ in the NE, $\textbf{F}_{-i}^{\text{NE}}$ is the combination of
precoding strategies of all users except user $i$ in the NE, and
$\textbf{F}_i^\prime$ stands for any possible precoding strategy
for user $i$.

In the cooperative scenario, all users are willing to cooperate
with each other and agree on a common principle in sharing the
resources. If the users choose the NB approach as a cooperation
principle, they aim at maximizing the Nash function (NF) defined
in the cooperative utility space (rate region) as \cite{Nash}
\begin{equation}\label{NF}
\mathfrak{F}=\prod\limits_{i\in \Omega_M} (R_i (\textbf{F}_i,
\textbf{F}_{-i}) - R_i^\prime)
\end{equation}
where $R_i^\prime$ is the information rate (the utility) that user
$i$ can achieve in the predefined \textit{disagreement point}
which the users will resort to if the cooperation breaks up.

In the NB game, the users need to specify also a manner of
cooperation according to which the bargaining is performed. It is
required that a particular manner of cooperation results in a
convex utility space. In the literature, the users are assumed to
cooperate with each other using orthogonal signaling schemes such
as TDM for FFCs and joint TDM/FDM for FSFCs
\cite{Leshem}--\cite{Leshem4}. It allows no interference among the
users. The main technical reason for considering orthogonal
signaling is that the rate region of a general interference
channel is yet unknown. Moreover, the use of orthogonal signaling
allows for power-efficient simple receives, while it is indeed
reasonable to assume that the users are equipped with simple
matched-filter-based receivers. In addition, if the users are
allowed to interfere with each other, they need to exchange the
interference information to achieve a desirable performance. It
may significantly increase the overhead in the system as well as
it also significantly complicates the transceiver design.
Therefore, orthogonal signaling is indeed a reasonable choice
which is also adopted here. It is worth mentioning, however, that
orthogonal signaling may be inefficient when the interference
among the users is low \cite{Zengmao}. In this case, the resulted
rate region may be a small subset of the actual rate region.
However, it is proved in \cite{Zengmao} that the cooperative
bargaining problem becomes convex even without orthogonal
signaling when the interference among users is small (as compared
to the channel noise), which renders the problem simpler in this
case. Moreover, the NE solution has a satisfactory performance in
the low-interference situation, and cooperation may not be the
best choice in this case considering the price paid for
coordinating the users \cite{Gesualdo}, \cite{Gesualdo2}.
Therefore, we focus on the case of high-interference in which
orthogonal signaling schemes are efficient, i.e., the cooperative
solutions based on orthogonal signaling achieves better
performance than the non-cooperative solutions.

It has been shown in \cite{Gesualdo} that the non-cooperative
precoding game can be simplified to a power allocation game under
orthogonal signaling. The following theorem shows that the
cooperative precoding game can be also simplified to a resource
allocation game if orthogonal signaling is used, i.e., TDM is used
in the case of FFCs or joint TDM/FDM is used in the case of FSFCs.

\textbf{Theorem 1}: {\it If the cooperation among users is based
on orthogonal signaling, the precoding matrix of each user in the
cooperative precoding game (\ref{e:Sysmodelgame}) which maximize
the NF (\ref{NF}) under the constraints (\ref{e:smc}) and
optionally (\ref{e:tpc}) adopts a strictly diagonal structure.}

\textbf{Proof}: See Appendix A.

Theorem $1$ can be interpreted as follows. In order to maximize
the utilities of all users in the cooperative precoding game, the
users must adjust their precoding matrices to achieve the
following two tasks: (i)~coordinating the utilization of frequency
bins (the public resources); (ii)~allocating powers (the
individual user resources) across the frequency bins. Therefore,
the cooperative precoding game is more complex compared to the
non-cooperative precoding game of \cite{Gesualdo} where the game
is solved by performing only individual power allocations among
frequency bins.

For further developments, two general assumptions need to be made:
(i)~The channel information of the desired channel
$\textbf{H}_{ii}$ is known at both the transmitter and receiver
sides of user $i$ only; (ii)~The TPCs are tight when they are
taken into account, i.e., $P^{\max}_i < \sum\limits_{k \in
\Omega_N} p_i^{\max}(k)$.

\vspace{-0.2cm}
\section{Cooperative Precoding/Resource Allocation Games with SMCs}
\label{s:smc} The following NB precoding/resource allocation
problem with only SMCs is considered
\begin{equation} \label{NBproblem1}
\mathop{\textbf{max}}\limits_{{\textbf{F}_i}, \, \forall i \in
\Omega_M} \; \prod\limits_{i\in \Omega_M} (R_i (\textbf{F}_i,
\textbf{F}_{-i}) - R_i^\prime) \quad \textbf{subject \,to:}  \;
[\textbf{F}_i \textbf{F}_i^H]_{kk} \leq {p_i^{\max}(k)}, \;
\forall i \in \Omega_M, \, \forall k \in \Omega_N .
\end{equation}

\vspace{-0.4cm}
\subsection{Cooperative strategies for two-user game}\label{s:fsc2}
The cooperative NB precoding/recourse allocation game
(\ref{NBproblem1}) is first considered for two-users only, i.e.,
$M = 2$ and $\Omega_M = \Omega_2$. Any stable point in the utility
region can be selected as a disagreement point. Since the NE point
given by \cite{Gesualdo}
\begin{equation} \label{e:NEfsCsmpf}
\textbf{F}_i^{\text{NE}}=\sqrt{{\rm diag}(\textbf{p}_i^{\max})},
\; \forall i \in \Omega_2
\end{equation}
is stable, it can be selected as a disagreement point. In
(\ref{e:NEfsCsmpf}), $\textbf{p}_i^{\max} = [p_i^{\max}(1),
\hdots, p_i^{\max}(N)]^T$ is the $N \times 1$ vector of power
limits on different frequency bins, i.e., the spectral mask
vector, and $( \cdot )^T$ and ${\rm diag} ( \cdot )$ stand for the
transpose and the operator that forms a square diagonal matrix by
writing the elements of a vector in the main diagonal,
respectively. It can be seen from (\ref{e:NEfsCsmpf}) that each
user exploits maximum allowed power on all frequency bins to
maximize its rate.

Knowing the disagreement point, the manner of cooperation between
users can be chosen as the joint TDM/FDM for FSFCs (see the
arguments in the previous section). The joint TDM/FDM prescribes
that any frequency bin can be used by only one user at any time
instant, but it may be shared by different users throughout the
operation time. The joint TDM/FDM can be implemented with low
complexity and the corresponding rate region is guaranteed to be
convex.

The following theorem about the structure of the optimal precoding
matrices of the two-user TDM/FDM cooperative game
(\ref{NBproblem1}) on FSFCs is in order.

\textbf{Theorem~2}: {\it The NB-based precoding/resource
allocation optimal strategies for the two-user TDM/FDM-based
cooperative precoding/resource allocation game (\ref{NBproblem1})
on the FSFCs are obtained through time sharing of at most two sets
of diagonal precoding matrices denoted as
${\{\textbf{F}_1^1,\textbf{F}_2^1\}}$ and
${\{\textbf{F}_1^2,\textbf{F}_2^2\}}$. The following conditions
must be satisfied }
\begin{equation}\label{e:ftimeshare2}
[\textbf{F}_i^l ]_{kk} \in \{0,\sqrt{p_i^{\max}(k)}\}, \, \forall i, \; \forall
k, \, \forall l; \quad \textbf{F}_1^l\textbf{F}_2^l
= \textbf{0}, \; \forall l; \quad \left({\rm Tr}\{\textbf{F}_i^1 -
\textbf{F}_i^2\}\right)^2 \in \mathfrak{P}_i, \, \forall i
\end{equation}
{\it where $i\in\Omega_2$, $l\in\Omega_2$, $k\in\Omega_N$, and
$\mathfrak{P}_i=\{{p_i^{\max}(k)} \, | \, k \in \Omega_N\}$ is the
set of power limits for user~$i$.}

\textbf{Proof}: See Appendix~B.

Theorem~2 states that the joint TDM/FDM-based cooperation on $N$
frequency bins can be realized by the time sharing of two diagonal
precoding matrices under SMCs. It can be also seen from the proof
of Theorem~2 (see Appendix~B) that only one frequency bin needs to
be shared. Denote this frequency bin as $k^\star$ and assume that
$[\textbf{F}_1^1 ]_{k^\star k^\star} = \sqrt{p_1^{\max}
(k^\star)}$. Therefore, $[\textbf{F}_1^2]_{k^\star k^\star} =
[\textbf{F}_2^1]_{k^\star k^\star} = 0$ and $[
\textbf{F}_2^2]_{k^\star k^\star} = \sqrt{p_2^{\max} (k^\star)}$.
Assuming that user~$1$ shares frequency bin~$k^\star$ for $\alpha$
portion of time ($0 \leq \alpha \leq 1$), the information rate of
user $i \in \Omega_2$ can be written as
\begin{eqnarray}\label{e:ratematrix}
R_i \!=\! \alpha \log \left( \det \left( \textbf{I} + \frac{(
{\textbf{F}^1_{i}})^{H} \mathbf{\Phi}_{ii}^{H} \mathbf{\Phi}_{ii}
\textbf{F}_{i}^1} {\sigma_i^2} \right) \right) \!+\! (1 - \alpha)
\log \left (\det \left( \textbf{I} + \frac{({ \textbf{F}_{i}^2}
)^H \mathbf{\Phi}_{ii}^{H} \mathbf{\Phi}_{ii} \textbf{F}_{i}^2}
{\sigma_i^2} \right) \right).
\end{eqnarray}

Therefore, the two-user cooperative NB precoding/recourse
allocation game (\ref{NBproblem1}) with the joint TDM/FDM
cooperation scheme can be converted to the problem of finding
$k^\star$ and $\alpha$ and be simplified as follows. The information
rate for user $i$ given in (\ref{e:ratematrix}) is the summation
of user $i$'s information rates on all frequency bins, and thus,
can be rewritten as
\begin{equation}\label{e:ratefscnb2}
R_i= \sum\limits_{k \in \Omega_N} \alpha_i (k) R_i(k),  \quad
\forall i \in \Omega_2
\end{equation}
where ${R_i(k) = \log (1 + |\mathbf{\Phi}_{ii}(k)|^2 p_i^{\max}(k)
/ \sigma_i ^2)}$ is the information rate that user~$i$ obtains
on frequency bin~$k$ by using it exclusively for all times, and
$\alpha_i (k)$ is the time portion during which the frequency
bin~$k$ is allocated to user~$i$. Note that in the NB solution of
the game, $0 < \alpha_i (k) < 1$ ($i \in \Omega_2$) hold only for
$k = k^\star$. Then, taking the logarithm of the NF, the NB
solution can be found by solving the following convex optimization
problem
\begin{equation} \label{e:NBmodel12}
\begin{split}
& \mathop{\textbf{max}}\limits_{\alpha_i (k), \, i \in \Omega_M, k
\in \Omega_N } \; \log(R_{1} - R_1^{\text{NE}}) + \log
(R_{2}-R_2^{\text{NE}}) \\
& \textbf{subject \,to:}  \quad \;\;\;  0\leq\alpha_i (k)\leq 1,
\; \forall i \in \Omega_2, \, \forall k \in \Omega_N  \\
& \quad\quad\quad\quad\quad\quad\;\;\; \alpha_1 (k)+\alpha_2 (k)
\leq1, \; \forall k \in \Omega_N; \quad R_{i} > R_i^{\text{NE}},
\; \forall i \in \Omega_2
\end{split}
\end{equation}
where $R_i^{\text{NE}}$ is the rate that user~$i$ obtains based
on the NE solution. It is worth noting that the last
constraint in (\ref{e:NBmodel12}) guarantees that both users can
achieve higher rates than $R_i^{\text{NE}}$ ($i \in \Omega_2$)
through the joint TDM/FDM-based cooperation. Otherwise the users
resort to the disagreement point and the cooperation breaks up.

\vspace{-0.3cm}
\subsection{Cooperative strategies for $M$-user game}\label{s:fscM}
Unlike the two-user case, where the NB solution of the cooperative
precoding/resource allocation game can be formulated as a time
sharing between two sets of precoding matrices, it is much more
complex to coordinate the users' precoding matrices in the
$M$-user game. The structure used for the two-user game can not be
directly applied here, especially when the number of users is
large. Therefore, to solve the $M$-user game, we first partition
time into time slots each of length $T$ to make it easier for the
users to perform time sharing. Moreover, considering the case when
the number of users or the channel states change over time, the
time partitioning enables a timely update of the bargaining
solution as long as time slots are small enough. In this case, the
cooperative solution can be obtained through the procedure
summarized in Table~\ref{table1}. In the following we focus on the
second step of the procedure in Table~\ref{table1}.

\begin{table}
\begin{center}
\caption {Procedure for finding cooperative solution in an $M$-user game.}
\label{table1}
\begin{tabular}{l}
\hline\hline \small{1. Initialization: the NE solution for the
precoding matrices is obtained and the NE point is used as a} \\
\small{disagreement point.} \\
\hline \small{2. Computation: the cooperative NB solution
for the precoding matrices is calculated.} \\
\hline \small{3. Implementation: Implement the NB solution for one
time slot. If any changes of the number of users or the} \\
\small{channel states are detected during this time slot, go back
to step~$1$ in the next slot; otherwise, repeat step~3.}
\\ \hline \hline
\end{tabular}
\end{center}
\vspace{-1.2cm}
\end{table}

As an extension of Theorem~2, the following theorem is in order.

\textbf{Theorem~3}: {\it Precoding matrices corresponding to the
NB solution of the TDM/FDM-based $M$-user cooperative game on the
FSFCs have the form }
\begin{equation}\label{e:nbfsc3}
\textbf{F}_i = \mathbf{\Gamma}_i(t) \sqrt { {\rm diag}
(\textbf{p}_i^{\max} )}, \; \forall i \in \Omega_M
\end{equation}
{\it where $\mathbf{\Gamma}_i(t)$ is a diagonal matrix with its
$k$th diagonal element }
\begin{equation}\label{e:U_i}
[\mathbf{\Gamma}_i (t)]_{kk} = \left\{ \begin{array}{ll}
1, & \textrm{if $t \in [b_i (k), e_i (k)]$} \\
0, & \textrm{if $t \notin [b_i (k), e_i (k)]$}\\
\end{array} \right.
\end{equation}
{\it with $b_i (k)$ and $e_i (k)$ representing, respectively, the
starting and ending time moments between which frequency bin~$k$
is allocated to user~$i$ in a time slot $[0, T]$. The following
conditions are then satisfied }
\begin{equation}\label{e:tscondition}
\begin{gathered}
\sum\limits_{i \in \Omega_M} \mathbf{\Gamma}_i(t)=\textbf{I};
\quad \mathbf{\Gamma}_i(t) \mathbf{\Gamma}_j(t) = \textbf{0}, \;
\forall i, j \in \Omega_M, \, j\neq i
\end{gathered}
\end{equation}
{\it where $t \in [0,T]$ is the time instant in a current time
slot. }

The proof of Theorem~3 is similar to the proof of Theorem~2 and is
omitted here. It is worth noting, however, that the difference is
that unlike the two-user game in which at most one frequency bin
needs to be shared, different groups of users may share different
frequency bins in the $M$-user game. The first condition in
(\ref{e:tscondition}) states that no frequency bin should be
vacant at any time, while the second condition in
(\ref{e:tscondition}) requests that no frequency bin be used by
more than one user at any time. Moreover, it is the length of
$[b_i (k), e_i  (k)]$, denoted as $\alpha_i (k) = e_i  (k) - b_i
(k)$, rather than the specific values of $b_i (k)$ and $e_i  (k)$,
that affects the rates of the users. Once the time portions
$\alpha_i (k)\; (\forall i\in \Omega_M, \forall k\in \Omega_N)$
are fixed, the order of using frequency bins is not important to
the users. Thus, the key problem is to calculate the time portions
$\alpha_i (k)\; (\forall i \in \Omega_M, \, \forall k \in
\Omega_M)$ that user~$i$ obtains on a frequency bin~$k$.
Mathematically, this problem can be formulated as the following
optimization problem
\begin{equation} \label{e:NBmodelM}
\begin{split}
& \mathop{\textbf{max}}\limits_{ \alpha_i (k), \, i \in \Omega_M,
k \in \Omega_N } \;\; \sum \limits_{i \in \Omega_M}
\log (R_i  - R_i^{\text{NE}}) \\
& \textbf{subject \,to:} \quad\quad \; 0\leq\alpha_i (k)\leq 1,
\; \forall i\in\Omega_M,\, \forall k\in\Omega_N \\
& \quad\quad\quad\quad\quad\quad \;\; \sum\limits_{i \in \Omega_M}
\alpha_i (k)\leq1, \; \forall k \in \Omega_N; \quad  R_i  >
R_i^{\text{NE}}, \, \forall i \in \Omega_M
\end{split}
\end{equation}
where $R_i$ is the sum of information rates that user~$i$
obtains on all frequency bins, that is,
\begin{equation}\label{e:rateperfbm}
R_i = \sum\limits_{k \in \Omega_N} \log \left( 1 + \frac{|
\mathbf{\Phi}_{ii} (k) \textbf{F}_{ii}(k) |^2}{\sigma_i^2} \right)
= \sum\limits_{k \in \Omega_N } \alpha_i (k) \log \left( 1 +
\frac{|\mathbf{\Phi}_{ii} (k)|^2 p_i^{\max}(k)}{\sigma_i^2}
\right) .
\end{equation}

To avoid a centralized channel estimation and information exchange
overhead among users on the cooperation stage, a
distributed algorithm for solving (\ref{e:NBmodelM}) is developed next.

\vspace{-0.3cm}
\subsection{Distributed algorithm for finding the NB solution}\label{s:da}
The problem (\ref{e:NBmodelM}) is a convex optimization problem
with a coupling constraint. Therefore, it can be solved in a
distributed manner using the dual decomposition method.

The Lagrange dual problem to (\ref{e:NBmodelM}) is
given as
\begin{equation} \label{e:lagrangian}
\begin{split}
& \mathop{\textbf{max}}\limits_{\alpha_i (k), \, i \in \Omega_M, k
\in \Omega_N } \;\; \sum\limits_{i \in \Omega_M} \log(R_i -
R_i^{\text{NE}}) - \sum\limits_{k \in \Omega_N} \lambda
(k) \left( \sum\limits_{i \in \Omega_M} \alpha_i (k)-1 \right) \\
& \textbf{subject \,to:} \quad\quad \; 0 \leq \alpha_i (k) \leq
1, \; \forall i \in \Omega_M, \, \forall k \in \Omega_N \\
& \quad\quad\quad\quad\quad\quad\;\;\;  R_i  > R_i^{\text{NE}}, \;
\forall i \in \Omega_M; \quad \lambda (k) \geq 0, \; \forall k \in
\Omega_N
\end{split}
\end{equation}
where $\lambda(k)\;(\forall k\in \Omega_N)$ are the positive
Lagrange multipliers.

The problem (\ref{e:lagrangian}) can be further converted into a
two-level optimization problem with the following lower level
subproblems
\begin{equation} \label{e:duallower}
\begin{split}
& \mathop{\textbf{max}}\limits_{\alpha_i (k), \, k
\in \Omega_N } \;\; \log(R_i  - R_i^{\text{NE}}) - \sum\limits_{k
\in \Omega_N} \lambda (k) \alpha_i (k) \\
& \textbf{subject \,to:} \quad\quad \; 0\leq\alpha_i (k) \leq 1,\;
\forall k \in \Omega_N, \quad  R_i > R_i^{\text{NE}}
\end{split}
\end{equation}
for each user $i \in \Omega_M$, and the higher level master
problem
\begin{equation}\label{e:dualhigher}
\mathop{\textbf{min}}\limits_{\lambda (k), \, k \in \Omega_N } \;
\sum\limits_{i \in \Omega_M} U_i (\boldsymbol{\lambda} ) +
\sum\limits_{k \in \Omega_N} \lambda (k) \quad \textbf{subject
\,to:} \; \lambda (k) \geq 0,\; \forall k \in \Omega_N
\end{equation}
where $U_i(\boldsymbol{\lambda})$ is the maximum value of the
objective function in (\ref{e:duallower}) given ${\boldsymbol
\lambda} = [\lambda (1), \hdots,$ $\lambda (N)]^T$.

The dual problem (\ref{e:duallower})--(\ref{e:dualhigher}) can be
solved based on a distributed structure with a coordinator. Since
the original problem is convex, strong duality holds and the
solutions of the dual problem (\ref{e:lagrangian}) and the
original problem (\ref{e:NBmodelM}) are the same if Slater's
condition is satisfied \cite{Covexoptmz}. For our specific
problem, we have the following result.

\textbf{Theorem~4}: {\it The Slater's condition is guaranteed to
be satisfied for the problem (\ref{e:NBmodelM}) as long as the NB
solution exists.}

\textbf{Proof}: See Appendix B.

Theorem~4 can be used to further simplify the lower level problem
(\ref{e:duallower}). Substituting (\ref{e:rateperfbm}) into the
objective function of the sub-problem (\ref{e:duallower}), the
latter can be rewritten as
\begin{equation} \label{e:duallower2}
\begin{split}
& \mathop{\textbf{max}}\limits_{\alpha_i (k), \,  k
\in \Omega_N } \;\; \log \left( \sum\limits_{k \in \Omega_N}
\alpha_i (k) R_i (k) - R_i^{\text{NE}} \right) - \sum\limits_{k
\in \Omega_N} \lambda (k) \alpha_i (k) \\
& \textbf{subject \,to:} \quad\quad \; 0 \leq \alpha_i (k) \leq
1,\; \forall k \in \Omega_N; \quad \sum\limits_{k \in \Omega_N}
\alpha_i (k) R_i (k) > R_i^{\text{NE}}
\end{split}
\end{equation}
where $R_i (k)={\log}(1+ |\mathbf{\Phi}_{ii}(k)|^2p^{\max}_i(k)/
\sigma_i ^2)$ is the rate on frequency bin~$k$ for user~$i$.
The lower level subproblems are solved distributively by the
corresponding users.

The Hessian of the objective function of the problem
(\ref{e:duallower2}) can be written as
\begin{equation}
\nabla^2 f_i(\boldsymbol{\alpha}_i) = - \left( \sum \limits_{k \in
\Omega_N} \alpha_i (k) R_i (k) - R_i^{\text{NE}}\right)^{-2}
\textbf{r}\textbf{r}^T
\end{equation}
where $\textbf{r} = [R_1 (1), \hdots, R_1 (N), R_2 (1), \hdots,
R_2 (N), \hdots, R_M (1), \hdots, R_M (N)]^T$ and
$\boldsymbol{\alpha}_i=[\alpha_i (1),...,$ $\alpha_i (N)]^T$. It
is straightforward to see that
$\nabla^2f_i(\boldsymbol{\alpha}_i)$ is negative definite since
$R_i (k) > 0$ ($\forall i \in \Omega_M, \forall k \in \Omega_N$).
Thus, each Lagrange problem (\ref{e:duallower2}) is guaranteed to
be strictly convex and a unique solution exists. More importantly,
the information required for solving the $i$th subproblem, i.e.,
$R_i (k)$ and $R_i^{\text{NE}}$, is local to user~$i$.

A coordinator is needed to solve the higher level master problem.
Since the overhead of the information exchange and the amount of
computations for (\ref{e:dualhigher}) is insignificant, any user
can act as a coordinator or all users can serve as coordinators in
a round-robin manner. The algorithm for solving the dual problem
is summarized in Table~\ref{table:dualalg}. Then, the complexity
of finding the bargaining solution is determined by the complexity
of the lower level subproblems (\ref{e:duallower2}) which is
$O(N^3)$.

\begin{table}
\begin{center}
\caption {Dual decomposition algorithm for NB.}
\label{table:dualalg}
\begin{tabular}{l}
\hline\hline
\small{1. The coordinator initializes $\boldsymbol{\lambda} =
\boldsymbol{\lambda}^0 = [\lambda^0(1), \lambda^0(2), ...,\lambda^0(N)]^T$
and broadcasts it to all users.} \\
\hline \small{2. Each user solves (\ref{e:duallower2}) according to the
present value of $\boldsymbol{\lambda}$ and transmits its solutions for} \\
\small{$\alpha_i (k), k \in \Omega_N$ to the coordinator.} \\
\hline \small{3. The coordinator updates $\boldsymbol{\lambda}$ according
to the gradient of the master problem (\ref{e:dualhigher}) as ${\hat \lambda
(k)} = [\lambda (k)$} \\
\small{$- \delta (1 - \sum\limits_{i \in \Omega_M} \alpha_i (k))
]_+$ ($\forall k \in \Omega_N$) where $(\cdot)_+$ denotes the projection
onto non-negative sub-space} \\
\small{and $\delta$ is the step length of the algorithm.}\\
\hline \small{4. If $|{{\hat\lambda (k)}-\lambda (k)|\leq\xi}
\,(\forall k\in \Omega_N)$, stop; otherwise the coordinator
broadcasts ${\boldsymbol{\hat{\lambda}}}$ and go to step~2.} \\
\small{Here, $\xi$ is the stopping threshold of the algorithm.}
\\ \hline \hline
\end{tabular}
\end{center}
\vspace{-1.2cm}
\end{table}

Note that the coefficients $\lambda(k)$ ($k \in \Omega_N$) have
specific physical meaning. Indeed, $\lambda(k)$ represents the risk
that cooperation among users breaks up due to a conflict on
sharing frequency bin~$k$. Thus, in the lower level subproblems,
the objective for each user consists of two parts. On one hand, a
larger $\alpha_i (k)$ is preferred to increase the total
information rate of user~$i$. On the other hand, if $\alpha_i (k)$
becomes too large, the cooperation may break up and the utility of
user~$i$ will return to the inferior competitive solution.

\vspace{-0.4cm}
\section{Cooperative Precoding/resource allocation games with SMCs
and TPCs}\label{s:smctpc} The following NB precoding/resource
allocation problem with both SMCs and TPCs is considered
\begin{equation} \label{NBproblem2}
\begin{split}
& \mathop{\textbf{max}}\limits_{\textbf{F}_i, \, \forall i \in
\Omega_M} \quad \;\; \prod\limits_{i\in \Omega_M} (R_i
(\textbf{F}_i, \textbf{F}_{-i}) - R_i^\prime) \\
& \textbf{subject \,to:}  \;\; [\textbf{F}_i \textbf{F}_i^H]_{kk}
\leq {p_i^{\max}(k)}, \; \forall i \in \Omega_M, \, \forall k \in
\Omega_N;  \quad {\rm Tr}\{\textbf{F}_i\textbf{F}_i^H\} \leq
P^{\max}_i, \; \forall i \in \Omega_M .
\end{split}
\end{equation}
Unlike the problem (\ref{NBproblem1}) considered in the previous
section, the diagonal elements of the precoding matrices
$\textbf{F}_i$ ($\forall i \in \Omega_M$) in (\ref{NBproblem2}) do
not necessarily satisfy
$[\textbf{F}_i]_{k,k}=\sqrt{p_{\text{max}}(k)}$ when frequency bin
$k$ is allocated to user $i$ because of the total power
constraint. However, if the joint TDM/FDM cooperation scheme is
used, Theorem~1 applies, and $\textbf{F}_i$ ($\forall i \in
\Omega_M$) can be written as
\begin{equation}
\textbf{F}_i={\rm diag}(\sqrt{\textbf{p}_i}), \, \forall i \in
\Omega_M.
\end{equation}

Using the same train of arguments as in the previous section,
(\ref{NBproblem2}) can be simplified as
\begin{equation}\label{e:NBmodelNC}
\begin{split}
& \mathop{\textbf{max}}\limits_{\alpha_i (k), \textbf{p}_i, \, i
\in \Omega_M, k \in \Omega_N } \; \sum\limits_{i \in
\Omega_M} \log (R_i - R_i^\prime) \\
& \textbf{subject \,to:} \; 0 \!\leq\! \alpha_i (k) \!\leq\! 1, \,
\forall i \in \Omega_M, \, \forall k \in \Omega_N; \;
\sum\limits_{i \in \Omega_M} \alpha_i (k) \!\leq\! 1, \, \forall k
\in \Omega_N; \; R_i  \!>\! R_i^\prime, \, \forall i \in \Omega_M \\
& \quad\quad\quad\quad\quad  \sum\limits_{k \in \Omega_N} \alpha_i
(k) p_i(k) \leq P^{\max}_i, \, \forall i \in \Omega_M; \quad
p_i(k) \leq p^{\max}_i(k), \; \forall i \in \Omega_M, \, \forall k
\in \Omega_N
\end{split}
\end{equation}
where $\textbf{p}_i=[p_i(1), p_i(2), \hdots, p_i(N)]$ is the power
allocation vector for user $i$, $R_i^\prime$ is the disagreement
point for user $i$, and $R_i = \sum\limits_{k \in \Omega_N} \log (
1 + |\mathbf{\Phi}_{ii}(k) \textbf{F}_{ii}(k) |^2 /
\sigma_i^2)=\sum\limits_{k \in \Omega_N} \alpha_i (k){\log}(1+
|\mathbf{\Phi}_{ii}(k)|^2 p_i(k) / \sigma_i^2)$ is the sum
information rate that user $i$ can obtain.

Unlike the problem (\ref{e:NBmodelM}) in the previous section, it
can be seen that $\textbf{p}_i$ ($i \in \Omega_M$) are also
optimization variables in (\ref{e:NBmodelNC}). Moreover,
(\ref{e:NBmodelNC}) is non-convex. Indeed, the Hessian matrix
$\textbf{H}_{f_i}$ of $f_i(\boldsymbol{\alpha}_i, \textbf{p}_i) =
\sum\limits_{k \in \Omega_M} \alpha_i (k) \textbf{p}_i (k)$ can be
written as
\begin{equation}
\textbf{H}_{f_i}=\nabla^2f_i(\boldsymbol{\alpha}_i,
\textbf{p}_i)=\left( {\begin{array}{*{20}c}
 {\textbf{0}} & {\textbf{I}} \\
 {\textbf{I}} & {\textbf{0}}\\
\end{array} } \right).
\end{equation}
Thus, $\textbf{H}_{f_i}$ ($\forall i \in \Omega_M)$ are orthogonal
matrices, i.e., $\textbf{H}_{f_i} \textbf{H}_{f_i}^T = \textbf{I}$
$(\forall i \in \Omega_M$). The eigenvalues of the orthogonal
matrices can only be $1$ or $-1$. Moreover, it is known that the
summation of all eigenvalues of $\textbf{H}_{f_i}$ equals ${\rm
Tr}\{\textbf{H}_{f_i}\}$ which is zero for any $i$. Therefore,
$\textbf{H}_{f_i}$ ($\forall i \in \Omega_M$) must have equal
number of eigenvalues $1$ and $-1$. The latter means that
$\textbf{H}_{f_i}$ ($\forall i \in \Omega_M$) are indefinite.
Thus, the constraints $\sum\limits_{k \in \Omega_N} \alpha_i (k)
p_i (k) \leq P^{\max}_i$ ($\forall i \in \Omega_M$) are
non-convex and the non-convexity of the optimization problem
(\ref{e:NBmodelNC}) follows.

In the following studies, the two-user case is considered and the
disagreement point is chosen at the origin of the rate region,
i.e., $R_i^\prime=0\,(\forall i \in \Omega_M)$, instead of the NE
point since finding the NE solution, in this case, is itself a
complicated problem.

\vspace{-0.3cm}
\subsection{Bandwidth-dominant and power-dominant systems}
\label{s:sysclass} Finding the TDM/FDM-based NB solution of the
problem (\ref{e:NBmodelNC}) requires joint power and frequency bin
allocation for each user, and the resulting complexity of the
two-user game can be unacceptably high. Moreover, the
TDM/FDM-based cooperation can be inefficient in some cases when
TPCs are present. To overcome these problems, we categorize
systems into two types and deal with each type separately. Toward
this end, two definitions need to be given first.

\textbf{Definition~1}: A point $\textbf{x}$ is {\it Pareto-optimal
} in space $\mathcal{S}$ if and only if $\textbf{y} = \textbf{x}$
for all $\textbf{y}$ satisfying $\textbf{y} \succeq \textbf{x}$ in
$\mathcal{S}$.

A Pareto-optimal point corresponds to an efficient allocation of
system resources. The NB solution is one of the Pareto-optimal
points in a utility space (rate region).

For the two-user cooperative game, there \ is \ a \ well \ known \
algorithm \ for \ obtaining \ the \ TDM/FDM-based NB solution if
only SMCs are imposed on the users \cite{Leshem}, \cite{angel1}.
According to this algorithm, the frequency bins are first arranged
such that $R_1(k)/R_2(k) \geq R_1(j)/R_2(j)$ ($\forall j, k \in
\Omega_N)$ if $k < j$, where $R_i(k)$ is the rate that
user~$i$ can achieve on frequency bin~$k$ by using
$p_i^{\text{max}} (k)$ and allowing no interference from
other users. Given any integer $\hat{k} \in \Omega_N$, let
\begin{equation}\label{POset}
\begin{split}
& \alpha_1 (k) = 1, \; \alpha_2 (k) = 0, \; p_1 (k) =
p_1^{\text{max}} (k), \; p_2(k) = 0, \; \forall k < \hat{k} \\
& \alpha_1 (k) = 0, \; \alpha_2 (k) = 1, \; p_1(k) = 0,\; p_2(k) =
p_2^{\text{max}} (k), \; \forall k > \hat{k} \\
& \alpha_1 (\hat{k}) = \beta, \; \alpha_2 (\hat{k}) = 1 - \beta,
\; p_1 (\hat{k}) = p_1^{\text{max}} (\hat{k}), \; p_2(\hat{k}) =
p_2^{\text{max}} (\hat{k}).
\end{split}
\end{equation}
Then the point $\textbf{R}=[R_1, R_2]$ is guaranteed to be
Pareto-optimal in the rate region for any $0 \leq \beta \leq 1$.
Varying $\hat{k}$ and $\beta$, all Pareto-optimal points can be
obtained including the NB solution of the game with only SMCs.

\textbf{Definition~2}: All Pareto-optimal points in a convex space
$\mathcal{S}$ form the \emph{Pareto-boundary} of $\mathcal{S}$.

The NB solution for the two-user cooperative
precoding/resource allocation game with only SMCs can be found by
searching on the Pareto-boundary instead of the entire utility
space of the game. The algorithm of \cite{Leshem} is based on the principle
that frequency bins which are ``better'' for a certain user should
be allocated to this user prior to the other frequency bins which
are ``inferior''. However, this principle may fail and lead to
highly inefficient solutions if TPCs are also imposed.

Consider the following simple example. Assume that there are four
frequency bins and $[R_1 (k),$ $R_2 (k)]$ are $[0.5, 0.1]$ for
$k=1$, $[2, 1]$ for $k=2$, $[1, 3]$ for $k=3$, and $[0.3, 1]$ for
$k=4$. Also assume that $\textbf{p}_i^{\text{max}}=[1, 1, 1, 1]$
and $P_i^{\text{max}} = 1.5$ for both users. Then according to the
aforementioned principle, the following resource allocation can be
obtained $\alpha_1 (1) =\alpha_2 (4) =1, \alpha_1 (2) = \alpha_2
(3) = 0.5, \alpha_1 (3) = \alpha_1 (4) = \alpha_2 (1) = \alpha_2
(2) = 0$ and $p_1 (1) = p_1 (2) = p_2 (3) = p_2 (4) = 1, p_1 (3) =
p_1 (4) = p_2 (1) = p_2 (2) = 0$. Note that the TPCs
$\sum\limits_{k \in \Omega_N} \alpha_i (k) p_i (k) \leq
P^{\max}_i$ ($i \in \Omega_2$) are used to derive the TDM/FDM
coefficients $\alpha_1 (2) = ( P_1^{\text{max}} - p_1^{\text{max}}
(1) \alpha_1 (1)) / p_1^\text{max} (2) = 0.5$ and $\alpha_2 (3) =
( P_2^{\text{max}} - p_2^{\text{max}} (4) \alpha_2 (4) ) /
p_2^\text{max} (3) = 0.5$. The resulting rates are $R_1 = 1.5$ and
$R_2 = 2.5$, and the point ($1.5, 2.5$) is obviously not
Pareto-optimal. For example, the strategies according to which
frequency bin~$2$ is allocated to user~$1$ and frequency bin~$3$
is allocated to user~$2$ for the whole time provide higher rates
than the allocation performed according to the aforementioned
principle. It is because the principle in \cite{Leshem} considers only comparative
advantages between the users, but not the absolute advantages.

It follows from the above discussion that the presence of TPCs
renders a different bargaining problem since there is a need to
coordinate between the power allocation and the frequency bin
allocation. Therefore, a different approach has to be developed.
Toward this end, note that we can first consider the solutions for
the bargaining game with only SMCs (denoted as game
$\mathcal{G}1$), and then add TPCs to the game (denoted as game
$\mathcal{G}2$).

\textbf{Observation~1}: TPCs do not enlarge the utility space of
the game. The Pareto-optimal solutions for game $\mathcal{G}1$
are also Pareto-optimal for game $\mathcal{G}2$ if they are
achievable.

Denote the Pareto-boundary of the TDM/FDM utility space of game
$\mathcal{G}1$ as $\mathcal{P}$. Then, the following
proposition is in order.

\textbf{Theorem~5}: {\it Assume that the frequency bins are
ordered such that $R_1 (k) / R_2 (k) \geq R_1 (j) / R_2 (j)$ ($
k,j\in\Omega_N$) if $k < j$. A non-empty subset $\mathcal{P}$ can
be achieved in game $\mathcal{G}2$ under both SMCs and TPCs if and
only if there exist $1 \leq \tilde{k} \leq N$ and $0 \leq
\tilde{\alpha} \leq 1$ such that }
\begin{equation}\label{POcon2}
\frac{P_1^{\text{max}} - \sum\limits_{k=1}^{\tilde{k} - 1 }
p_1^{\text{max}} (k) }{p_1^{\text{max}} ({\tilde{k}})} \geq
\tilde{\alpha} \geq \frac{\sum\limits_{k = \tilde{k}}^N
p_2^{\text{max}} (k) - P_2^{\text{max}}} {p_2^{\text{max}}
(\tilde{k})}.
\end{equation}

\textbf{Proof}: See Appendix~C.

According to (\ref{POcon2}), all multi-user systems can be
categorized into bandwidth- and power-dominant. If
condition (\ref{POcon2}) is satisfied, the system is
bandwidth-dominant and the rates of both users can increase
simultaneously only if new frequency bins are added into the
system. Otherwise, the system is power-dominant and the rates of
both users can increase simultaneously only when TPCs of the users
are relaxed.

\textbf{Observation~2}: Beginning as a bandwidth-dominant, a
multi-user system gradually changes towards a power-dominant as the
number of available frequency bins increases.

\subsection{Bandwidth-dominant systems: TDM/FDM based bargaining}
In the bandwidth-dominant systems, the TDM/FDM-based cooperation
is efficient in the sense that a non-empty subset $\mathcal{P}$
can be achieved in game $\mathcal{G}2$. Denote the
Pareto-boundary of the TDM/FDM utility space of $\mathcal{G}2$ as
$\mathcal{P}_{2}$. Then, for the bandwidth-dominant systems, the
bargaining can be restricted in the set $\mathcal{P}^\prime =
\mathcal{P}_{2} \cap \mathcal{P}$ only. The resulted NB solution,
denoted as $S_{NB}^\prime$, can be sub-optimal as compared to the
optimal solution of the non-convex optimization problem
(\ref{e:NBmodelNC}). It is because the power allocation (which is
not the dominant factor in this case) is not optimized jointly
with frequency bins allocation.

Denote the optimal NB solution of game $\mathcal{G}2$ as
$S_{NB}^{opt}$. Also denote the TDM/FDM utility spaces of
games $\mathcal{G}1$ and $\mathcal{G}2$ as $\mathcal{U}_1$ and
$\mathcal{U}_2$, respectively. Then, the following theorem
regarding the optimality of $S_{NB}^\prime$ is in order.

\textbf{Theorem~6}: {\it $S_{NB}^\prime = S_{NB}^{opt}$ if
$S_{NB}^{opt} \in \mathcal{P}^\prime$. If $S_{NB}^\prime \neq
S_{NB}^{opt}$, then $S_{NB}^{opt} \notin \mathcal{P}$ but
$S_{NB}^\prime \in \mathcal{P}$, which means that $S_{NB}^{opt}$
is not Pareto-optimal in $\mathcal{U}_1$ but $S_{NB}^\prime$ is
Pareto-optimal in $\mathcal{U}_1$. }

\textbf{Proof}: See Appendix~C.

Theorem~6 leads to the following two conclusions about the
optimality of $S_{NB}^\prime$ in the bandwidth-dominant systems:
(i)~$S_{NB}^\prime$ can be identical to the optimal TDM/FDM based
NB solution; (ii)~$S_{NB}^\prime$ is guaranteed to be
Pareto-optimal in $\mathcal{U}_1$ (which is larger than
$\mathcal{U}_2$) even if the optimal NB solution is not
Pareto-optimal.

\subsection{Power-dominant systems: FDM/sampled time sharing-based bargaining}
Let us now consider the case of power-dominant systems. The example given in
Subsection~\ref{s:sysclass} is, in fact, an example of a power-dominant
system. From this example, we can make the following observation.

\textbf{Observation~3}: The use of the maximum allowed power on
all allocated frequency bins generally results in a non-optimal
solution for game $\mathcal{G}2$.

To verify this observation, let us denote the set of all frequency
bins as $\mathcal{B}$, the subset of frequency bins which user~$1$
occupies using the maximum allowed power as $\mathcal{B}_1^{\rm
max}$, the set of frequency bins which user~$2$ occupies using the
maximum allowed power as $\mathcal{B}_2^{\rm max}$. Then, user~1
may improve its rate by water-filling on $\mathcal{B} -
\mathcal{B}_2^{\rm max}$ instead of using the maximum allowed
power on $\mathcal{B}_1^{\rm max}$, while the rate of user~2 can
be kept the same. Here $\mathcal{B} - \mathcal{B}_2^{\rm max}$
denotes the difference between sets $\mathcal{B}$ and
$\mathcal{B}_2^{\rm max}$, and the general term {\it
water-filling} is used to represent the specific meaning of
finding the solution of the following convex problem
\begin{equation} \label{waterfil}
\begin{split}
& \mathop{\textbf{max}}\limits_{p_i(j), \, j \in \Omega_N} \;
\sum\limits_{j \in \Omega_N} \log (1 + \varepsilon_i(j) p_i(j) )
\quad \textbf{subject \,to:} \; \sum\limits_{j \in \Omega_N} p_i(j)
= P_i^{\max}; \; p_i(j) \leq p_i^{\text{max}} (j), \, \forall j \in \Omega_N
\end{split}
\end{equation}
which is a single-user multi-channel power allocation problem with
constant $\varepsilon_i(j) = |[\mathbf{\Phi}_{ii}]_{jj}|^2 / \sigma_i^2$
being a measure of the channel~$j$ for user $i$, which depends on
the channel gain and channel noise.

Observation~3 suggests that the power-dominant games have to be
played based on a different manner of cooperation from the
TDM/FDM. A reasonable choice of the manner of cooperation is the
FDM/time sharing (TS), which considers time sharing between points
corresponding to different FDM based frequency bin allocation
schemes. Then the power allocation, which is the dominant problem
in this case, is based on the water-filling problem
(\ref{waterfil}). However, the complexity of finding the FDM/TS
based NB solution is high, especially when the number of frequency
bins is large. To obtain the FDM/TS-based NB solution, the
water-filling should first be performed for all $2^N$ possible
frequency bin allocations between the users, and the resulted
$2^N$ points in the utility space should be recorded. Then the TS is used
to obtain a minimum convex space containing all these points and
the NB solution is derived. The complexity of the TS is then
$O(4^N)$, which is exponential in the number of frequency bins.

To reduce the complexity, we consider a simplified version of the
FDM/TS, which is the FDM/sampled time sharing (STS). The proposed
FDM/STS scheme finds the optimal FDM/STS based NB solution
according to the algorithm described in Table~\ref{poweralg}.

\begin{table}
\begin{center}
\caption {Algorithm for the power-dominant systems.}
\label{poweralg}
\begin{tabular*}{0.88\textwidth}[]{p{14.5cm}}
\hline\hline 1. Both users $i \in \Omega_2$ perform the
water-filling (\ref{waterfil}) on $\mathcal{B}$ and obtain two
sets of frequency bins $\tilde{\mathcal{B}_i}$ ($i \in \Omega_2$).
Then, $\mathcal{B}_c= \tilde{\mathcal{B}_1} \bigcap \tilde
{\mathcal{B}_2}$ is the set of frequency bins under competition.\\
\hline 2. In the first round of this step, user~$1$ is allocated
the set of frequency bins $\tilde{\mathcal{B}_1}$ and user~$2$
performs water-filling on $\mathcal{B}-\tilde{\mathcal{B}_1}$. In
round~$j$ ($j$ goes from $2$ to $L=|\mathcal{B}_c|$ where
$|\mathcal{B}_c|$ denotes the cardinality of the set
$\mathcal{B}_c$), user~$1$ selects a subset, denoted as
$\mathcal{B}_s^j$, of $j-1$ frequency bins with smallest channel
gains from the set $\mathcal{B}_c$ and performs water-filling on
$\mathcal{B}-\mathcal{B}_s^j$. Then, user~$2$ performs
water-filling on the remaining frequency bins.
After the $L$th round, $L$ points in the utility space are obtained. \\
\hline 3. Perform $L$ rounds of the aforementioned step~2 for
user~$2$ starting from the state that user $2$ is allocated the
set of frequency bins $\tilde{\mathcal{B}_2}$ and user~$1$
performs water-filling on $\mathcal{B}-\tilde{\mathcal{B}_2}$.
Obtain other $L$ points in the utility space.\\
\hline 4. Denote the set of $2L$ points obtained in steps~2~and~3
as $\mathcal{T}$. Find the Pareto-boundary $\mathcal{P_T}$ of
$\mathcal{S_T}$ where  $\mathcal{S_T}$ is the minimum convex space
containing $\mathcal{T}$.\\
\hline 5. Bargain on $\mathcal{P_T}$ and obtain the solution
$S_{NB}^{\prime \prime}$.\\
\hline \hline
\end{tabular*}
\end{center}
\vspace{-1.2cm}
\end{table}

Let $\mathcal{WF}^i(\mathcal{X})$ denotes the water-filling
operator for user~$i$ on the set of frequency bins $\mathcal{X}$.
It returns the maximum rate that user~$i$ can obtain by optimizing
its power allocation on $\mathcal{X}$. Let also the vector of
rates $\textbf{R}^{opt}$ corresponding to the FDM/TS-based NB
solution $S_{NB}^{opt}$ be obtained by time sharing of two points
$(R_1^{opt1}, \, R_2^{opt1} )$ and $(R_1^{opt2}, \, R_2^{opt2})$
in the utility space of game $\mathcal{G}2$, and the time
sharing coefficients are $\lambda$ and $1 - \lambda$,
respectively, that is, $\textbf{R}^{opt} = (\lambda R_1^{opt1} +
(1 - \lambda) R_1^{opt2}, \lambda R_2^{opt1} + (1 - \lambda)
R_2^{opt2})$. Denote the sets of frequency bins allocated to the
users in the points $(R_1^{opt1}, \, R_2^{opt1})$ and
$(R_1^{opt2}, \, R_2^{opt2})$ as $(\mathcal{B}_1^{opt1}, \,
\mathcal{B}_2^{opt1})$ and $(\mathcal{B}_1^{opt2}, \,
\mathcal{B}_2^{opt2})$, correspondingly. Then, the following
theorem is in order.

\textbf{Theorem~7}: {\it The FDM/STS based NB solution
$S_{NB}^{\prime \prime}$ obtained using the algorithm in
Table~\ref{poweralg} can be identical to the FDM/TS based NB
solution $S_{NB}^{opt}$. If they are not identical, the difference
$d$ between the logarithm of the NF for $S_{NB}^{opt}$ and the
logarithm of the NF for $S_{NB}^{\prime \prime}$ is bounded by }
\begin{equation}
d < \min \left( \log \left( \frac{\mathcal{WF}^2
(\mathcal{B}_2^{opt2})} {\mathcal{WF}^2 (\mathcal{B} -
\tilde{\mathcal{B}_2})} \right), \, \log \left(
\frac{\mathcal{WF}^1 (\mathcal{B}_1^{opt1})} {\mathcal{WF}^1
(\mathcal{B} - \tilde{\mathcal{B}_1})} \right) \right) .
\end{equation}

\textbf{Proof}: See Appendix~C.

The following conclusions can be drawn regarding $S_{NB}^{\prime
\prime }$ in the power-dominant systems: (i)~$S_{NB}^{\prime
\prime}$ is the optimal FDM/STS based NB solution. Thus, it is a
Pareto-optimal solution in the FDM/STS utility space;
(ii)~$S_{NB}^{\prime \prime}$ can be identical to $S_{NB}^{opt}$;
(iii)~The efficiency of $S_{NB}^{\prime
\prime}$ depends on the ratios $\omega_1$ and $\omega_2$, where
$\omega_i = \mathcal{WF}^i(\mathcal{B}_i^{opt}) / \mathcal{WF}^i
(\mathcal{B} - \tilde{\mathcal{B}_i})$.

\subsection{The two-user algorithm}
The overall algorithm, which combines both the bandwidth-dominant
and power-dominant cases, for the two-user cooperative NB game is
given in Table~\ref{alg}.
\begin{table}
\begin{center}
\caption {The overall algorithm for the two-user NB game with SMCs
and TPCs.}\label{alg}
\begin{tabular}{l}
\hline\hline \small{1. Check the condition (\ref{POcon2}):}
\small\emph{If it is satisfied, go to step~2},
\small\emph{otherwise, go to step~3.}\\
\hline \hline\small{2. System is bandwidth-dominant:}
\small\emph{Search on the Pareto-boundary
$\mathcal{P}^\prime$, and return the solution $S_{NB}^\prime$.}\\
\hline \small{3. System is power-dominant:} \small\emph{Derive
$\tilde{\mathcal{B}_1}$, $\tilde{\mathcal{B}_2}$, and
$\mathcal{B}_c$. Play the $2L$ rounds and obtain $\mathcal{T}$ and
$\mathcal{P_T}$.} \\
\small\emph{\quad Search on $\mathcal{P_T}$, and return the
solution $S_{NB}^{\prime \prime}$.} \\
\hline \hline
\end{tabular}
\end{center}
\vspace{-1.2cm}
\end{table}

In the bandwidth-dominant case, the complexity of searching on
$\mathcal{P}^\prime$ is $O(N)$. In the power-dominant case, the
complexity of the algorithm in Table~\ref{poweralg} is determined
by the time sharing part, which is $O(L^2)$, i.e., the complexity
reduction as compared to $O(4^N)$ for the optimal FDM/TS based
solution (where the time consumed on water-filling is neglected in
both cases) is dramatically significant, especially for large $N$.

\section{Simulation results}\label{s:sr}

\subsection{Cooperative precoding/resource allocation games with SMCs}
In the first example, we assume that two users share four
available frequency bins. The noise power $\sigma^2$ is $0.01$ for
both users on all frequency bins. The channel gains of the desired
channels $\mathbf{\Phi}_{11}$ and $\mathbf{\Phi}_{22}$ are
generated as Rayleigh random variables with mean $1$. The channel
gains of the interference channels $\mathbf{\Phi}_{12}$ and
$\mathbf{\Phi}_{21}$ are generated as Rayleigh random variables
with means $0.7$ and $0.2$, respectively. The elements of the
spectral mask vector $\textbf{p}_{\text{max}}$ are also Rayleigh
random variables with mean $1$.

\begin{figure}
\begin{center}
\includegraphics[angle=0,width=0.75\textwidth]{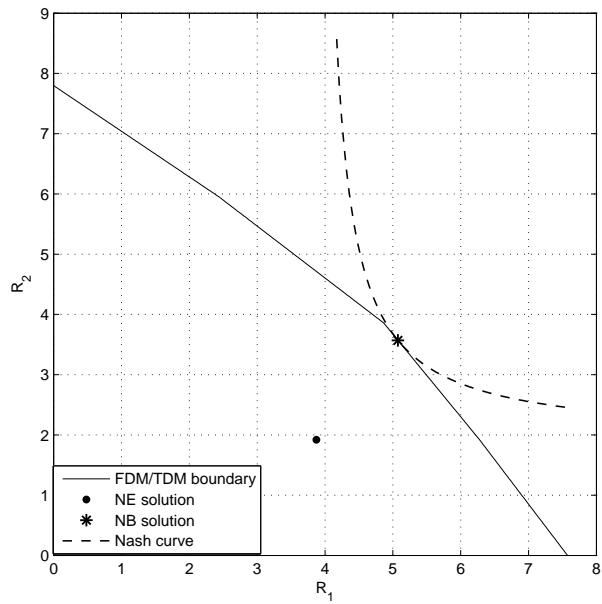}
\end{center}
\caption {The FDM/TDM rate region and the NE and NB solutions on
the frequency selective channel.} \label{FSC_rateregion}
\end{figure}

\begin{figure}
\begin{center}
\includegraphics[angle=0,width=0.7\textwidth]{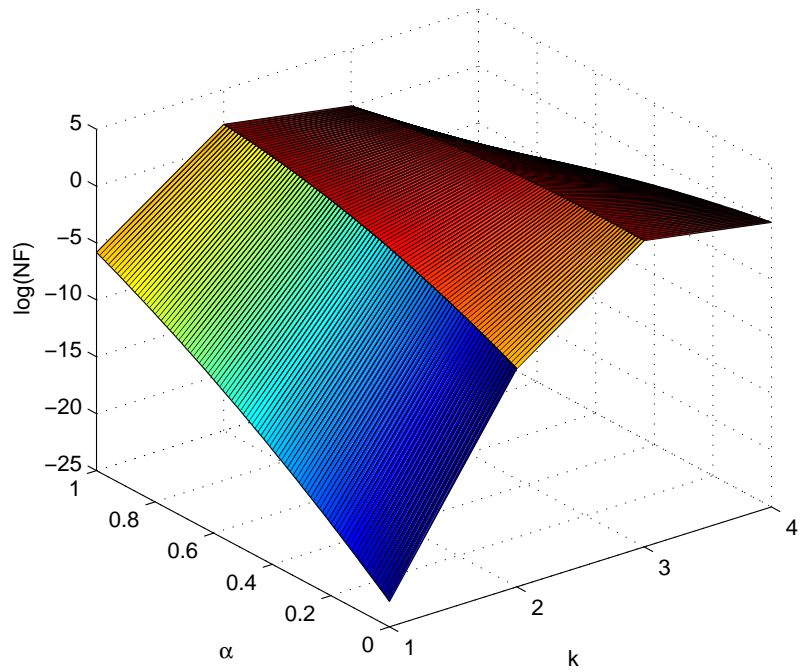}
\end{center}
\caption {The logarithm of the NF under different FDM/TDM
frequency bin allocation schemes.} \label{NFschemes}
\end{figure}

In Fig.~\ref{FSC_rateregion}, the NB solution computed according
to Theorem~2 is shown together with the NE solution. The boundary
of the TDM/FDM rate region is also included in the figure.
Fig.~\ref{NFschemes} displays the values of the logarithm of the NF under different
TDM/FDM frequency bin allocation schemes. In this figure, $k$ is
the frequency bin being shared and $\alpha$ is the fraction of
time that user~$1$ uses the frequency bin $k$. It can be seen in
Fig.~\ref{FSC_rateregion} that the NB solution lies on the
boundary of the TDM/FDM rate region and provides significantly
larger rates to both users than the NE solution. Moreover, the NB
solution is fair to both users. It can be also seen in
Fig.~\ref{NFschemes} that the largest value of the logarithm of the NF corresponds
to the optimal scheme that provides the NB solution.

In the second example, the distributed algorithm for the $M$-user
game developed in Section~\ref{s:da} is tested. It is assumed that
four users share six frequency bins. As in the previous example,
channel gains of the desired and interference channels are
generated as Rayleigh random variables with means 1 and 0.2,
respectively. The elements of the spectral mask vector
$\textbf{p}_{\text{max}}$ are also Rayleigh random variables with
mean $1$. The step length $\delta = 0.2$ (if different values are
not specified) and stopping threshold $\xi = 10^{-5}$ are
selected.

\begin{figure}
\begin{center}
\includegraphics[angle=0,width=0.7\textwidth]{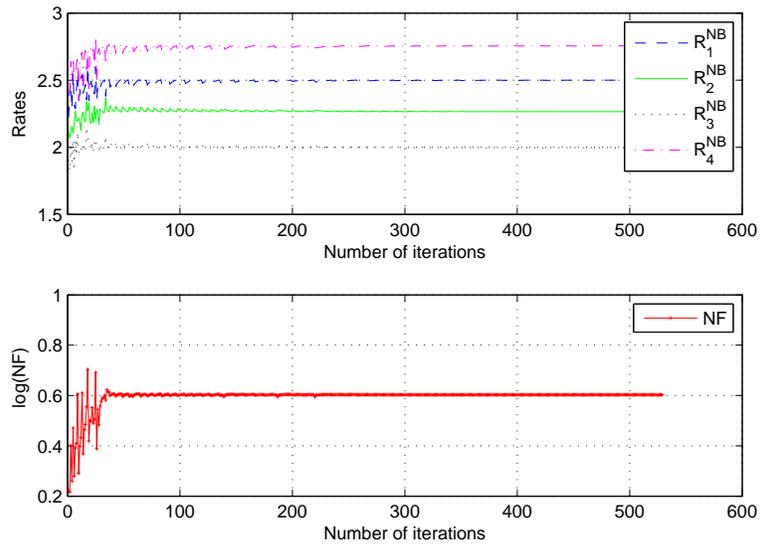}
\end{center}
\vspace{-0.4cm} \caption {Instantaneous information rates and the
corresponding logarithm of NF versus number of iterations.}
\label{NF&Rate}
\end{figure}

\begin{figure}[h]
\begin{center}
\includegraphics[angle=0,width=0.75\textwidth]{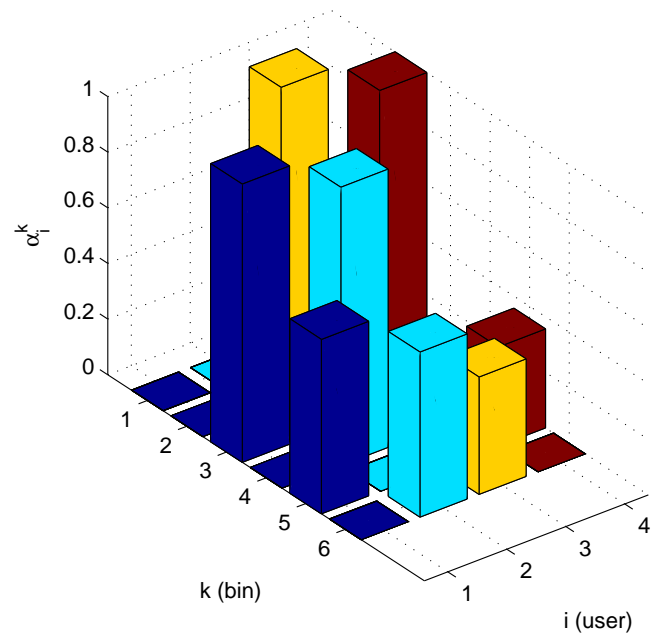}
\end{center}
\vspace{-0.4cm} \caption {Allocations of time portions on
frequency bins $\{\alpha_i (k)\}$.} \label{NF&alpha}
\end{figure}

The iterations of the NB process are shown in Fig.~\ref{NF&Rate}.
The four curves on the upper side of the figure show the
instantaneous information rates that the corresponding users can
achieve, and the curve at the bottom shows the corresponding values of
the logarithm of the NF. 
The NB and NE solutions and the comparison between them in terms
of the percentage of improvement provided by the NB solution
versus the NE solution are shown in Table~\ref{t:FSC4user} for one
of the runs. It can be seen from Fig.~\ref{NF&Rate} and
Table~\ref{t:FSC4user} that all users obtain supplementary benefit
from cooperation. The corresponding final allocation of time
portions on each frequency bin for each user is shown in
Fig.~\ref{NF&alpha}. It can be seen that frequency bins $1$, $2$,
$3$, and $4$ are occupied exclusively by users $3$, $4$, $1$, and
$2$, respectively, while frequency bins $5$ and $6$ are shared by
users $1$ and $4$, and users $2$ and $3$, respectively.

\begin{table}
\begin{center}
\caption {Comparisons between NE and NB}\label{t:FSC4user}
\begin{tabular}{r||c|c|c}\hline
User& NE Solution& NB solution & Increased by\\
\hline\hline 1 & 1.1296 & 2.2707 & 101.02\% \\
\hline 2 & 1.4014 & 2.4906 & 77.72\% \\
\hline 3 & 1.2952 & 2.3992 & 85.24\% \\
\hline 4 & 1.6957 & 2.4175 & 42.56\% \\ \hline
\end{tabular}
\end{center}
\vspace{-1.2cm}
\end{table}

\begin{figure}[h]
\begin{center}
\includegraphics[angle=0,width=0.7\textwidth]{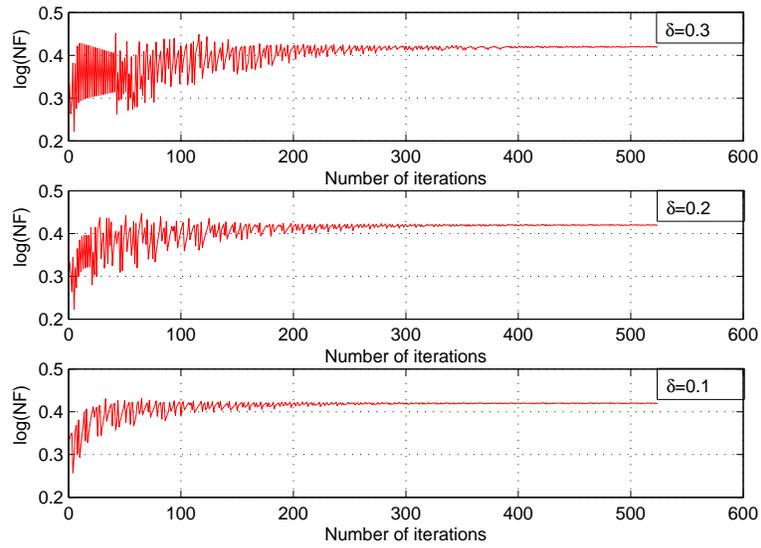}
\end{center}
\vspace{-0.4cm} \caption {The logarithm of the NF versus number of
iterations under different step lengths, $\delta\in\{ 0.3, 0.2,
0.1\}$.} \label{NF&steplength}
\end{figure}

Fig.~\ref{NF&steplength} depicts the effect of the step length on
the convergence speed of the algorithm. With the step lengths
$\delta\in\{ 0.1, 0.2, 0.3\}$, the corresponding logarithm of NF
is shown in each sub-figure. It can be seen that the algorithm is
time-efficient with a good choice of the step length.

\subsection{Cooperative precoding/resourse allocation games with
SMCs and TPCs} 
\begin{figure}
\begin{center}
\includegraphics[angle=0,width=0.75\textwidth]{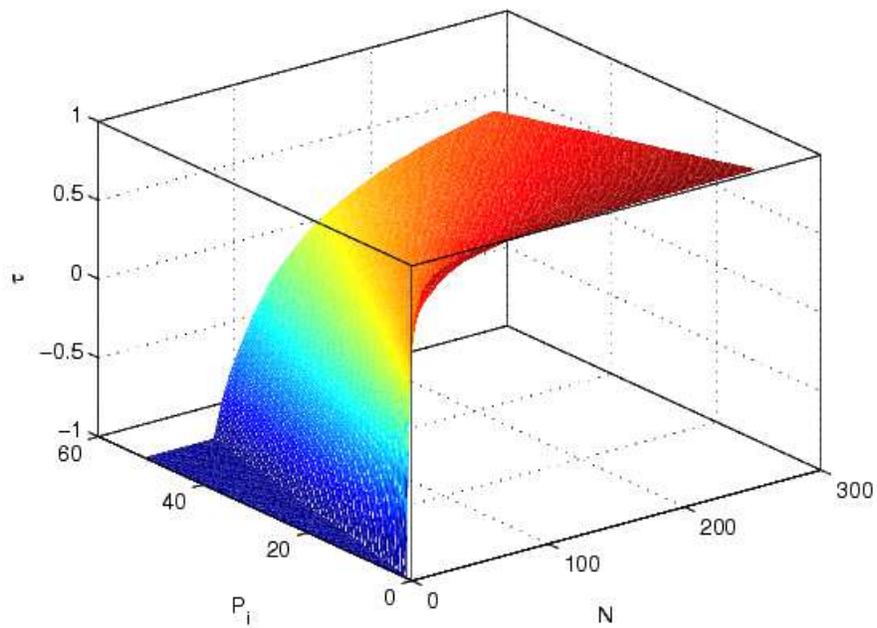}
\end{center}
\vspace{-0.6cm} \caption {System classification versus total power
limits and number of frequency bins.} \label{BD&PD}
\end{figure}

Fig.~\ref{BD&PD} shows the system classification
according to Theorem~5 versus the total power limits and the
number of frequency bins for the two-user system. The total power
limits of the users $P_1^{\max}$ and $P_2^{\max}$ are equal and
vary from $1$ to $51$. The number of frequency bins $N$ increases
from $1$ to $256$. The desired channel gains are randomly
generated using Rayleigh distribution with mean 1, and the users
do not interfere with each other due to the orthogonal signaling
assumption. The power limits on different frequency bins
$p^{\text{max}}_i (k)\;(\forall i \in \Omega_2, \, \forall k \in
\Omega_N)$ are uniformly distributed in the interval $[1.8, \,
2.2]$. The frequency bins are sorted such that $R_1 (k) / R_2 (k)
\geq R_1 (j) / R_2 (j)\;(k,j\in\Omega_N)$ if $k < j$. Following
the comparative advantage based principle introduced in
Section~\ref{s:sysclass}, the maximum number of frequency bins
$k_i$ that user $i$ can cover is $k_1=\{\max\,t_1|t_1\in
\Omega_N,\,\sum_{k=1}^{t_1}p_1^{\max}(k)\leq P_1\}$ for user 1 and
$k_2=\{ \max\,t_2|t_2\in
\Omega_N,\,\sum_{k=N-t_2+1}^{N}p_2^{\max}(k)\leq P_2\}$ for user
2. The total normalized bandwidth $b_i$ (with the bandwidth of
each frequency bin normalized to 1) that user $i$ can cover is
then $b_1 = k_1 + \left( P_1 - \sum_{k=1}^{k_1}p_1^{\max} (k)
\right) / p_1^{\max} (k_1+1)$ for user 1 and $b_2 = k_2 + \left(
P_2 - \sum_{k=N-k_2+1}^{N}p_2^{\max}(k)\right)/p_2^{\max}(N-k_2)$
for user 2. Then the variable $\tau =1- (b_1 + b_2) / N$ stands
for the system property characteristic according to Theorem~5. The
system is bandwidth-dominant if $-1 \leq \tau \leq 0$ and is
power-dominant if $0 < \tau < 1$. It can be seen from the figure
that the system changes gradually from bandwidth- to
power-dominant when new frequency bins are added into the system,
while it changes gradually from power- to bandwidth-dominant when
the total power limits of the users are relaxed.

In our last example, the power-dominant two-user system is
considered.\footnote{Note that for the bandwidth-dominant systems,
the algorithm for finding the NB solution inherits the algorithm
for finding the optimal TDM/FDM based NB solution in the precoding
games without TPCs which is already studied above.} The number of
frequency bins varies from $4$ to $9$ ($50$ runs for each case).
The total power limits of the users are set as $P_i = 2$ ($i \in
\Omega_2$) for each user, and the power limits on different
frequency bins are set to $1 + x (k)$ where $x(k)$ is a uniform
random variable in the interval $[0.2, \, 0.25]$. It guarantees
that the system is power-dominant. The channel gains on all
frequency bins are randomly generated for both users using
Rayleigh distribution with mean 1.

\begin{figure}
\begin{center}
\includegraphics[angle=0,width=0.75\textwidth]{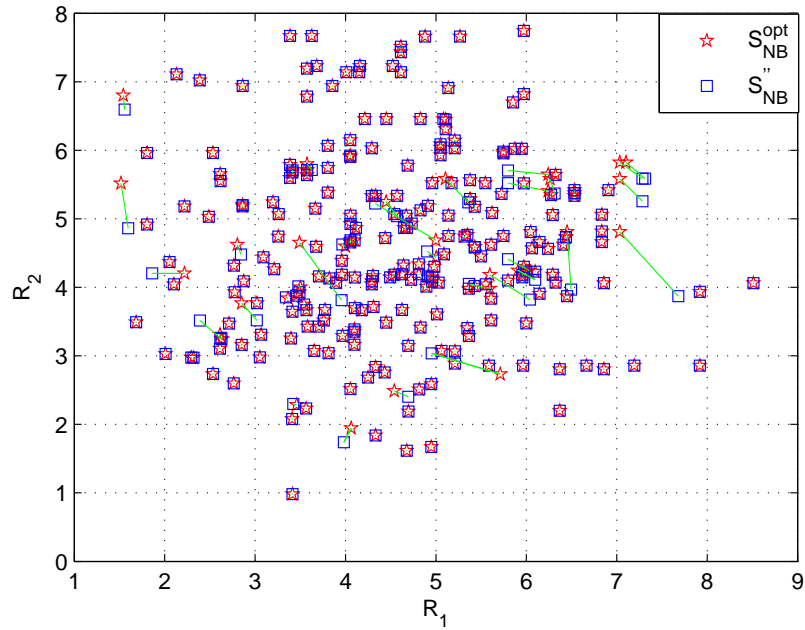}
\end{center}
\vspace{-0.6cm}
\caption {$S_{NB}^{opt}$ and $S_{NB}^{''}$ in the rate region.} \label{accuracy}
\end{figure}

\begin{figure}[h]
\begin{center}
\includegraphics[angle=0,width=0.75\textwidth]{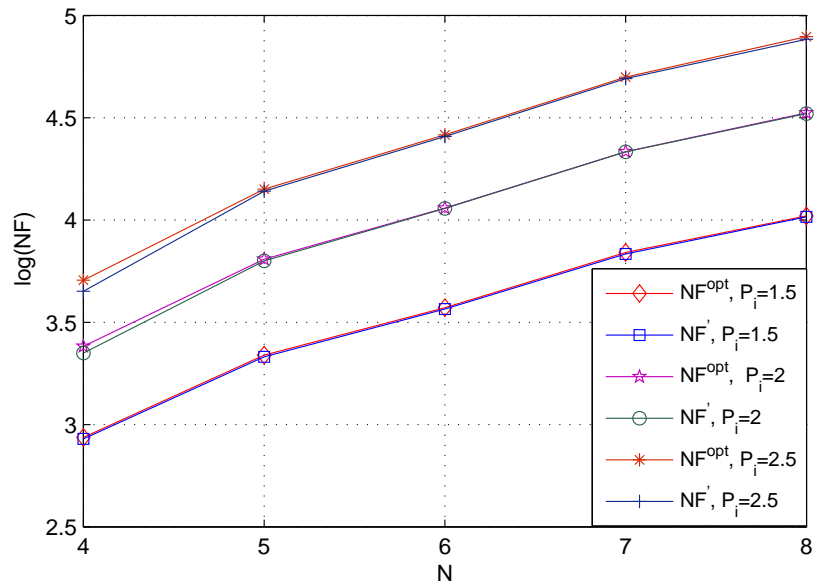}
\end{center}
\vspace{-0.6cm} \caption {The logarithm of the NF for the FDM/TS
and FDM/STS based NB solutions.} \label{NF_opt_sub_2}
\end{figure}

Fig.~\ref{accuracy} shows the FDM/TS- and FDM/STS-based NB
solutions $S_{NB}^{opt}$ and $S_{NB}^{\prime \prime}$,
respectively, in $300$ simulation runs. It can be seen in the
figure that $S_{NB}^{\prime \prime}$ is identical to
$S_{NB}^{opt}$ for most of the cases. Moreover, although the
distance between $S_{NB}^{\prime \prime}$ and $S_{NB}^{opt}$ for
some cases may appear relatively large in the utility space, the
difference between the values of the logarithm of their NF are
small as shown in Fig.~\ref{NF_opt_sub_2}. Particularly,
Fig.~\ref{NF_opt_sub_2} depicts the logarithm of the NF for
$S_{NB}^{opt}$ and $S_{NB}^{\prime \prime}$ (denoted as
$\text{NF}^{opt}$ and $\text{NF}^\prime$, respectively) versus the
number of frequency bins $N$ when the total power limits
$P_i^{\max}$ ($i \in \Omega_2$) are set to $1.5$, $2$, or $2.5$.
Every point in the figure is averaged over $50$ runs. It can be
seen from Fig.~\ref{NF_opt_sub_2} that the gap between
$\text{NF}^{opt}$ and $\text{NF}^\prime$ is very small, if it is
not zero.

\section{Conclusions}\label{s:con}
Cooperative NB-based precoding/resource allocation strategies on
FSFCs are studied under SMCs and optionally TPCs. First, it is
shown that the optimal precoding matrices adopt a strictly
diagonal structure and the NB-based precoding game is equivalent
to a corresponding resource allocation game if the users are not
allowed to interfere with each other, i.e., orthogonal signaling
is used. The use of orthogonal signaling is practically important
since it significantly simplifies transceiver design and allows
for significant reduction of the system overhead during user
cooperation. Second, it is shown that the NB solution of the
cooperative precoding/resource allocation game with only SMCs can
be obtained efficiently in a distributed manner (a simple
coordinator is required) with inevitable information exchanges
among users. The developed two-level user cooperation procedure
avoids a large system overhead by enabling users to perform most
of the computations individually using their local information.
Third, it is shown that the cooperative NB-based
precoding/resource allocation game with both SMCs and TPCs is
non-convex and finding its optimal solution requires joint
optimization of the frequency bins (which is the public resource)
and each user's transmit power (which is the individual resource)
allocations. The complexity of finding the optimal solution is
unacceptably high in this case. Therefore, it is proposed to
categorize all multi-user systems into bandwidth- and
power-dominant depending on the bottleneck resource in the system.
For different classes of the systems, the algorithms based on
different manners of cooperation are developed. While the TDM/FDM
based cooperation is still efficient for the bandwidth dominant
systems, the TDM/STS cooperation is used for the power-dominant
systems. The above classification of the multi-user systems
guarantees that the solutions obtained by the algorithms
corresponding to each category are Pareto-optimal and can be even
identical to the optimal solutions, while the algorithm complexity
is significantly reduced. Simulation results demonstrate the
effectiveness of the proposed cooperative solutions and their
superiority to the NE solutions. 

\section*{Appendix A: Proof of Theorem~1 in Section~\ref{s:sm_long}}
The noise covariance in (\ref{e:initem}) is equivalent to
$\textbf{R}_{-i}=\sigma_i^2\textbf{I}$ when the cooperation among
users is based on orthogonal signaling such as, for example, TDM
for FFCs or joint TDM/FDM for FSFCs.\footnote{It can be, however,
any other orthogonal signaling scheme.} Thus,
$R_i(\textbf{F}_i,\textbf{F}_{-i})$ is simplified to
\begin{equation}\label{ratesimple}
R_i = \log \left( \det \left( \textbf{I} + \frac{1}{\sigma_i^2}
\textbf{F}_{i}^{H} \mathbf{\Phi}_{ii}^{H} \mathbf{\Phi}_{ii}
\textbf{F}_{i} \right) \right).
\end{equation}
The Hadamard's inequality ($\det (\textbf{A} ) \leq \prod_i
a_{ii}$ for a Hermitian positive semidefinite matrix $\textbf{A}$)
suggests that the determinant in (\ref{ratesimple}) is maximized
when $\textbf{F}_{i}$ is diagonal. Moreover, the power constraints
given in (\ref{e:smc}) and (\ref{e:tpc}) are irrelevant to the
non-diagonal elements of $\textbf{F}_i$. Therefore, the optimal
precoding matrices must be diagonal. \hfill $\blacksquare $

\section*{Appendix B: Proof of Theorems~2~and~4 in Section~\ref{s:smc}}

\subsection*{Proof of Theorem 2}
As follows from Theorem~1, the optimal $\textbf{F}_1^l$ and
$\textbf{F}_2^l$ ($l \in \Omega_2$) are diagonal. The three
conditions in (\ref{e:ftimeshare2}) are based on the fact that the
joint TDM/FDM is used. First, consider the FDM part. Given any
division of the frequency bins, both users will use maximum
allowed power on all frequency bins allocated to them. Thus, the
first condition in (\ref{e:ftimeshare2}) follows.

The second condition in (\ref{e:ftimeshare2}) is based on the fact
that only one user is allowed on any given frequency bin at any
time. Thus, there must be a user which allocates zero power on any
given frequency bin at any time.

The third condition in (\ref{e:ftimeshare2}) is based on the fact
that the NB solution can be obtained by sharing at most a single
frequency bin between both users.\footnote{Note that a similar
observation has been made in \cite{Leshem2}, but here we give a
different much simpler proof.} The proof of the latter fact can be
given by contradiction using the optimality of the NB solution.

Assume that the NB solution can be obtained only by sharing two or
more frequency bins between both users and consider the case when
two frequency bins $m$ and $n$ are shared. In the sharing scheme,
user $2$ uses a fraction $\alpha_1$ of the time in frequency bin
$m$ and a fraction $\alpha_2$ of the time in frequency bin $n$.
Let $R_i (m)$ and $R_i (n)$ be the rates that user $i$ can obtain
by exclusively using frequency bins $m$ and $n$, respectively.
Without loss of generality, we assume that $R_2 (m)/R_1 (m) \geq
R_2 (n)/R_1 (n)$. Then either of the following cases must happen:
(i) if $\alpha_1 R_2 (m) + \alpha_2 R_2 (n) \ge R_2(m)$, there
exists $\gamma\in [0,1)$ such that $\alpha_1 R_2 (m) + \alpha_2
R_2 (n) = R_2 (m) + \gamma R_2 (n)$; (ii) if $\alpha_1 R_2 (m) +
\alpha_2 R_2 (n) < R_2 (m)$, there exists $\gamma \in [0,1)$ such
that $\alpha_1 R_2 (m) + \alpha_2 R_2 (n) = \gamma R_2 (m)$.

Case~(i) corresponds to the sharing scheme according to which only
frequency bin $n$ is shared between both users, and user~$2$
exploits a fraction $\gamma$ of time on frequency bin $n$.
According to this new sharing scheme, user~$2$ obtains the same
rate on frequency bins $m$ and $n$ as that in the original scheme.
Then the rate that user~$1$ can obtain on frequency bins $m$ and
$n$ in the new scheme is
\begin{eqnarray}
(1 \!-\!\gamma)R_1 (n) \!\!\!&=&\!\!\! \left( 1 \!-\!
\frac{(\alpha_1 \!-\! 1) R_2 (m) \!+\! \alpha_2 R_2 (n)}{R_2 (n)}
\right) R_1 (n) \!=\! (1 \!-\! \alpha_1) \frac{R_1 (n) R_2
(m)}{R_2 (n)} \!+\! (1 \!-\! \alpha_2) R_1 (n) \nonumber \\
\!\!\!&\geq&\!\!\! (1 \!-\! \alpha_1) R_1 (m) \!+\! (1 \!-\!
\alpha_2) R_1(n). \label{e:contra_proof_user1}
\end{eqnarray}
The last inequality follows from the assumption that ${R_2 (m)/R_1
(m) \geq R_2 (n)/R_1 (n)}$. It can be seen from
(\ref{e:contra_proof_user1}) that the rate that user~$1$ can
obtain using the new sharing scheme is equal to or larger than
that in the original scheme. This contradicts the assumption that
the NB solution can be achieved only by sharing of two frequency
bins between both users.

A similar result can be derived for Case~(ii). Moreover, when more
than two frequency bins are shared, the above proof can be used
iteratively to obtain the same result. Therefore, the optimal
solution can be obtained by sharing at most a single frequency bin
between both users. Thus, $\textbf{F}_i^1$ ($i \in \Omega_2$) can
be obtained by adding/deleting a single diagonal element of
$\textbf{F}_i^2$ ($i \in \Omega_2$) and the third condition in
(\ref{e:ftimeshare2}) follows. \hfill $\blacksquare $

\subsection*{Proof of Theorem 4}
Since the constraints of the problem (\ref{e:NBmodelM}) are all
linear, the Slater's condition reduces to two parts with the first
part requiring that the feasible domain of $f = \sum\limits_i \log
(R_i - R_i^{\text{NE}})$ be open and the second part requiring
that the feasible domain of the whole problem be non-empty.

It is straightforward to verify that the first part is satisfied.
The second part is equivalent to the requirement of the existence
of the NB solution. This completes the proof. \hfill $\blacksquare $

\section*{Appendix C: Proof of Theorems~5--7 in Section~\ref{s:smctpc}}

\subsection*{Proof of Theorem 5}
First note that in game $\mathcal{G}1$ any resource allocation
scheme satisfying (\ref{POset}) results in a Pareto-optimal point
in the utility space, and vice versa. Thus, the statement of the
theorem is equivalent to the statement that (\ref{POcon2}) is the
sufficient and necessary condition to guarantee that at least one
set of $\{\hat{k}, \beta \}$ satisfies the conditions
(\ref{POset}). To prove the sufficiency, let (\ref{POcon2}) is
satisfied, $\hat{k} = \tilde{k}$, and $\beta = \tilde{\alpha}$ in
(\ref{POset}). Then the resulting total powers used by the users
are $P_1^\prime = \sum\limits_{k=1} ^{\tilde {k} - 1}
p_1^{\text{max}}(k) + \tilde{\alpha}p_1 ^{\text{max}} (\tilde{k})$
for user $1$ and $P_2^\prime = \sum\limits_{k = \tilde{k}+1}^N
p_2^{\text{max}} (k) + (1 - \tilde{\alpha}) p_2^{\text{max}}
(\tilde{k})$ for user $2$. Using (\ref{POcon2}), it is easy to
verify that $P_1^{\text{max}} \geq P_1^\prime$ and
$P_2^{\text{max}} \geq P_2^\prime$. Therefore, the sufficiency is
proved. The necessity can be proved similarly using contradiction.
\hfill $\blacksquare $

\subsection*{Proof of Theorem 6}
The first part of the theorem follows from the \emph{independence
on irrelevant alternatives} property of the NB \cite{gtbook}. This
property states that bargaining in a convex subset which contains
the NB solution of the original set results in the same NB
solution. Thus, it is clear that if $S_{NB}^\prime \neq
S_{NB}^{opt}$, then $S_{NB}^{opt} \notin \mathcal{P}^\prime$.
Since $\mathcal{P}^\prime$ is the achievable subset of
$\mathcal{P}$ in game $\mathcal{G}2$, it is impossible that
$S_{NB}^{opt} \in \mathcal{P}$ and $S_{NB}^{opt} \notin
\mathcal{P}^\prime$ simultaneously. Thus, if $S_{NB}^{opt} \notin
\mathcal{P}^\prime$, then $S_{NB}^{opt} \notin \mathcal{P}$ as
well. This completes the proof. \hfill $\blacksquare $

\subsection*{Proof of Theorem 7}
Let $R_1^{opt1} > R_1^{opt2}$. Then $R_2^{opt1} < R_2^{opt2}$ due
to the Pareto-optimality. Let also $\textbf{R}^1 = (R_1^{1},
R_2^{1})$ and $\textbf{R}^2 = (R_1^{2},R_2^{2})$ be two points
generated in steps~2 and 3 of the algorithm summarized in
Table~\ref{poweralg} such that $R_1^{1} \geq R_1^{opt1}$ and
$R_2^{2} \geq R_2^{opt2}$. Denote the sets of frequency bins
allocated to the users in the points $\textbf{R}^1$ and
$\textbf{R}^2$ as $(\mathcal{B}_1^{1}, \mathcal{B}_2^{1})$ and
$(\mathcal{B}_1^{2}, \mathcal{B}_2^{2})$, respectively. Recalling
that $R_i^\prime=0$ ($\forall i \in \Omega_2$), i.e., the
disagreement point is the origin, the difference between the
logarithm of the NF for $\textbf{R}^{opt}$ and the logarithm of
the NF for $\textbf{R}^1$ can be obtained as
\begin{eqnarray}
d_1 \!\!\!&=&\!\!\! \log \left( \lambda R_1^{opt1} \!+\! (1 \!-\!
\lambda) R_1^{opt2} \right) \!+\! \log \left( \lambda R_2^{opt1}
\!+\! (1 \!-\! \lambda) R_2^{opt2} \right) \!-\! \log (R_1^{1})
\!-\! \log (R_2^{1}) \nonumber \\
\!\!\!&=&\!\!\! \log \left( \lambda \frac{R_1^{opt1}} {R_1^{1}} +
(1 - \lambda) \frac{R_1^{opt2}} {R_1^{1}} \right) + \log \left(
\lambda \frac{R_2^{opt1}} {R_2^{1}} + (1 - \lambda ) \frac{R_2^
{opt2}} {R_2^{1}} \right) \nonumber \\
\!\!\!&<&\!\!\! \log \left( \lambda \frac{R_1^{opt1}} {R_1^{1}} +
(1 - \lambda ) \frac{R_1^{opt1}} {R_1^{1}}\right) + \log \left(
\lambda \frac{R_2^{opt2}} {R_2^{1}} + (1 - \lambda )
\frac{R_2^{opt2}}{R_2^{1}}\right) \nonumber \\
\!\!\!&\leq&\!\!\! \log \left( \lambda \frac{R_2^{opt2}} {R_2^{1}}
+ (1 - \lambda ) \frac{R_2^{opt2}} {R_2^{1}} \right) = \log \left(
\frac{R_2^{opt2}} {R_2^{1}} \right) = \log \left( \frac{
\mathcal{WF}^2 (\mathcal{B}_2^{opt2})} { \mathcal{WF}^2 (
\mathcal{B} - \mathcal{B}_1^{j}) } \right) . \label{d1der}
\end{eqnarray}
where the inequalities hold because $\log \left( \lambda
R_1^{opt1} + (1 - \lambda ) R_1^{opt2} \right) + \log \left(
\lambda R_2^{opt1} + (1 - \lambda )R_2^{opt2} \right) - \log
(R_1^{1}) - \log (R_2^{1}) > 0$, $R_1^{opt1} > R_1^{opt2},
R_2^{opt1} < R_2^{opt2}$, and $R_1^{1}>R_1^{opt1}$, and the last
equality is obtained by substituting the notations $R_2^{opt2} =
\mathcal{WF}^2 (\mathcal{B}_2^{opt2})$ and $R_2^{1} =
\mathcal{WF}^2 (\mathcal{B} - \mathcal{B}_1^{j})$. Here $1 \leq j
\leq L$ stands for the index of the round in which $R_2^{1}$ is
obtained by user~$2$.

Furthermore, using the fact that $\mathcal{WF}^2 ( \mathcal{B} -
\mathcal{B}_1^{j} ) \geq \mathcal{WF}^2 (\mathcal{B} -
\tilde{\mathcal{B}_2})$, (\ref{d1der}) can be simplified as
\begin{equation}
d_1 < \log \left( \frac{\mathcal{WF}^2 (\mathcal{B}_2^{opt2})}
{\mathcal{WF}^2 (\mathcal{B} - \tilde{\mathcal{B}_2})} \right).
\end{equation}

It can be derived in a similar way that the difference $d_2$
between the logarithm of the NF for $\textbf{R}^{opt}$ and the
logarithm of the NF for $\textbf{R}^2$ obeys the following
inequality
\begin{equation}
d_2 < \log \left( \frac{\mathcal{WF}^1 (\mathcal{B}_1^{opt1})}
{\mathcal{WF}^1 (\mathcal{B} - \tilde{\mathcal{B}_1})} \right).
\end{equation}

Finally, note that neither $\textbf{R}^1$ nor $\textbf{R}^2$ have
been assumed to be the rates corresponding to the optimal
FDM/STS-based NB solution. Indeed, $\textbf{R}^1$ and
$\textbf{R}^2$ are just two of $2L$ points generated in steps~2
and 3 of the algorithm summarized in Table~\ref{poweralg},
respectively. Thus, the rates corresponding to the actual solution
returned by the algorithm are expected to be superior to the rates
corresponding to the points $\textbf{R}^1$ and $\textbf{R}^2$, or
even equal to the rates in the optimal solution
$\textbf{R}^{opt}$.\footnote{As an example, when
$|\mathcal{B}_c|=0$, the solution obtained by the algorithm in
Table~\ref{poweralg} is identical to the optimal solution
$\textbf{R}^{opt}$.} Therefore, $d \leq \min(d_1, d_2)$ and $d$
can be equal to zero. This completes the proof. \hfill
$\blacksquare $

\newpage

\newpage

\end{document}